\documentclass[11pt]{article}
\usepackage{amsmath}
\usepackage{hyperref}
\usepackage{graphicx}
\usepackage{bm}
\usepackage{xcolor}
\usepackage{amssymb}
\usepackage{amsfonts}
\usepackage{comment}
\usepackage{cite}
\usepackage{todonotes}
\usepackage{caption}
\usepackage{subcaption}
\usepackage{CJKutf8}
\usepackage{authblk}
\usepackage{diagbox} 
\usepackage[utf8]{inputenc}

\usepackage{mathrsfs}
\def\be{\begin{equation}}
\def\ee{\end{equation}}
\def\ba{\begin{eqnarray}}
\def\ea{\end{eqnarray}}
\def\nn{\nonumber}

\def\bl#1\el{\begin{align}#1\end{align}}

\title{ Primordial Black Hole Formation in Dust-Radiation Bouncing Cosmologies}

 \date{}

\def\be{\begin{equation}}
\def\ee{\end{equation}}
\def\ba{\begin{eqnarray}}
\def\ea{\end{eqnarray}}
\def\nn{\nonumber}
\def\bl#1\el{\begin{align}#1\end{align}}

\large

\newcommand{\dpar}[1]{\left(#1 \right)} 
\newcommand{\dcol}[1]{\left[#1 \right]} 
\newcommand{\dcha}[1]{\left\{#1 \right\}} 
\usepackage{xcolor}
\usepackage{cancel}

\usepackage{esvect}
\usepackage{enumitem}

\def\be{\begin{equation}}
\def\ee{\end{equation}}
\def\ba{\begin{eqnarray}}
\def\ea{\end{eqnarray}}
\def\nn{\nonumber}

\def\bl#1\el{\begin{align}#1\end{align}}

\author[a,b]{Xuan Ye\thanks{ yexuan@cuhk.edu.cn}}
\author[c]{ Luiz Felipe Demétrio\thanks{demetrio.luizfelipe.fis@gmail.com }}
\author[d]{Eduardo José Barroso\thanks{eduardojose.barroso@lapp.in2p3.fr}}
\author[e,f]{Shen-Feng Yan\thanks{sfyan@cdut.edu.cn }}
\author[g]{Nelson Pinto-Neto\thanks{nelsonpn@cbpf.br}}
\affil[a]{\small School of Science and Engineering, The Chinese University of Hong Kong,
Shenzhen, 518172, Guangdong,China}
\affil[b]{\small Department of  Astronomy,  Key Laboratory
for Researches in Galaxies and Cosmology,
 University of Science and Technology of China,  Hefei, Anhui, 230026,  China}
\affil[c]{\small Departamento de Física, Universidade Estadual de Londrina,
Rod. Celso Garcia Cid, Km 380, 86057-970, Londrina, Paraná, Brazil}
\affil[d]{\small Université de Savoie, CNRS, IN2P3, LAPP, Annecy-le-Vieux, France}
\affil[e]{\small College of Physics, Chengdu University of Technology, Chengdu 610059, China}
\affil[f]{\small School of Physics, The University of Electronic Science and Technology of China, 88 Tian-run Road, Chengdu, China}
\affil[g]{\small COSMO– Centro Brasileiro de Pesquisas Físicas,
Rua Dr. Xavier Sigaud 150, 22290-180, Rio de Janeiro– RJ, Brasil}

\begin{document}
\begin{CJK}{UTF8}{gbsn}
\maketitle

\allowdisplaybreaks

\begin{abstract}

Primordial black holes (PBHs) provide a unique probe of the early Universe and may have an enhanced abundance in bouncing cosmologies, where a long contracting phase can amplify perturbations. We develop a unified framework to study PBH formation in dust–radiation bouncing cosmologies, focusing on the classical contracting phase so that the results are insensitive to bounce details.
We compute the curvature power spectrum for an extremely small dust equation of state using a stable semi-analytical (adiabatic) method, derive  the Jeans length of the two‑fluid system
using  dynamical‑system analysis and the WKB approximation, and extend the three-zone model from the single- to the two-fluid case to model local collapse. We implement two collapse criteria to obtain the curvature perturbation threshold for PBH formation and estimate PBH mass fractions for benchmark masses spanning low-mass ($10^{-17} M_{\odot}$)  to supermassive ($10^{13} M_{\odot}$) scales. The critical curvature threshold is extremely small and nearly mass-independent over a broad range $(\zeta_c \sim 10^{-21}$ for $10^{-14}$ to $10^{13} M_{\odot})$, with deviations only near dust–radiation equality. Nevertheless, the square root of the curvature power spectrum at the relevant formation times is many orders of magnitude smaller, yielding vanishingly small PBH mass fractions across the benchmark masses. Compared with the pure-dust case, radiation pressure and the two-fluid collapse conditions significantly suppress PBH production, implying that substantial PBH formation in dust–radiation bouncing cosmologies would require additional mechanisms to amplify curvature perturbations.

\end{abstract}

\section{Introduction}\label{sec1}

Primordial black holes (PBHs), which form through non-stellar processes, were first proposed in the early 1970s~\cite{zel1966hypothesis, hawking1971gravitationally, carr1974black, carr1975primordial}. They are believed to form in the early universe from the gravitational collapse of overdense regions 
that exceed a critical threshold~\cite{harada2013threshold, musco2019threshold, escriva2020universal}. Since their formation occurs before Big Bang nucleosynthesis~\cite{cyburt2003primordial}, PBHs are non-baryonic and thus constitute a viable Dark Matter (DM) candidate~\cite{chapline1975cosmological, clesse2018seven, carr2020primordial,Khlopov:2010,Belotsky:2014,Belotsky:2019,sym16111487}. Their formation mechanism allows for an extremely broad mass spectrum, ranging from the Planck mass to supermassive black hole scales \cite{Choptuik:1992jv, Niemeyer:1999ak, Shibata:1999zs, Harada:2023ffo, Musco:2004ak, Musco:2012au, Harada:2013epa, Sadhukhan:2021chg, Yoo:2024lhp, Ye:2025wif, Franciolini:2018vbk, Pi:2017gih, Chen:2020uhe, Carr:2020gox, Cai:2018tuh, Cai:2021wzd, Cai:2023ptf, Chen:2022usd, Chen:2016kjx,haonan}, although only PBHs with masses above roughly $10^{15} \mathrm{g}$ would survive to the present day without being fully evaporated by Hawking radiation~\cite{hawking1975particle}. However, very small-mass PBHs can also play an important role in the early universe; for example, they may potentially solve the monopole and domain wall problems. Depending on their formation mass, the present PBH abundance in DM can be constrained by comparing their expected signatures with a variety of observational probes, allowing PBHs to constitute either a fraction or, in some cases, the entirety of DM~\cite{carr2021constraints, villanueva2021brief}.

The majority of PBH formation studies have been conducted within the context of inflationary cosmology~\cite{bullock1997non, yokoyama1998chaotic, motohashi2017primordial, martin2020primordial, clesse2015massive}, where PBHs are expected to form when perturbations re-enter the horizon during the radiation dominated stage, and PBH formation can occur via various mechanisms \cite{Saini:2017tsz}.  
In such scenarios, only two narrow mass windows, approximately $10^{-17} M_{\odot} \sim 10^{-11}M_{\odot}$ and $10^{13} M_{\odot} \sim 10^{19} M_{\odot}$, allow PBHs with a monochromatic mass function to account for all of the DM, as reviewed in~Ref.\cite{Carr2025}. To explore a broader spectrum of PBH masses, non-inflationary cosmologies are attractive for investigating, bouncing cosmology is the most popular scenario where the long contracting phase and decreasing scale factor may enhance scalar perturbation modes for producing PBH and gravitational waves~\cite{Barroso2025, carr2011persistence, papanikolaou2024primordial, quintin2016black, banerjee2022primordial,CHEN2017561,Cai:2023uhc,Zhu:2024ffj}.


In Ref.~\cite{Barroso2025}, we investigated  PBH formation in a pure-dust bouncing cosmology by analyzing the spherical collapse of matter perturbations. In that work, perturbations characterized by a density contrast $\delta$ were assumed to undergo gravitational collapse and form PBHs whenever they exceeded a critical threshold $\delta > \delta_c$. The main task was therefore to establish an appropriate prescription for determining  $\delta_c$. Because the equation of state is dust-like, the collapse dynamics can be accurately described by the Lemaître–Tolman–Bondi (LTB) class of solutions. Evaluating these solutions under the assumption that the maximum collapse radius is set by the Hubble length, which decreases during the contracting phase, leads to a time-dependent critical density threshold. Within this framework, the resulting PBH abundance was computed and found to be negligible for PBH masses that would survive to the present epoch. Since only dust was present and no additional denser fluids were considered, the mass spectrum was dominated by low-mass PBHs that would have evaporated by today. Consequently, the overall PBH abundance in that scenario was concluded to be cosmologically irrelevant.

 As a natural extension, we study PBH formation in more realistic bouncing cosmologies that include both dust and radiation. For completeness, we consider the two-fluid configuration of radiation and dust as a particular case of Ref.\cite{vitenti2013quantum}, in which both the background and its linear perturbations are quantized consistently through the Wheeler-De Witt equation without ever invoking the classical background equations. This framework allows one to track the evolution of cosmological fluctuations, including those capable of seeding PBH formation through the entire bouncing phase in a self-consistent quantum setting. Since PBH formation is studied in the classical contracting phase, our conclusions on PBH formation are universal for dust-radiation bouncing models and insensitive to the details of the quantum bouncing phase.

For the two‑fluid model, we follow an analysis similar to that of Ref.~\cite{Barroso2025}, but several key elements must be modified. The presence of radiation prevents the use of exact analytic LTB-type solutions for the collapse, making a numerical or semi‑analytical treatment necessary. As a result, the prescription for the critical threshold employed in the pure-dust case cannot be  applied directly. Moreover, including radiation introduces extra coupled equations of motion of each fluid component. These must be integrated together with the quantum field fluctuation modes for various momenta, a task that much more complex than the computational framework used in the previous work. The present study therefore adapts and extends the methodology to accommodate the richer dynamics of the two-fluid system.

Following Refs. \cite{Barroso2025, carr1974black}, we adopt the Press–Schechter formalism to estimate the mass fraction of PBHs for four representative masses at their formation time during the contracting phase. The paper is  thus naturally divided into three parts. In the first part, we investigate the perturbation theory of the two‑fluid quantum bouncing model during contraction. We derive the power spectrum of curvature perturbations for the four wavenumbers corresponding to the masses of interest, which serve as the variance of curvature perturbations under a Gaussian distribution. The adiabatic \footnotemark{} method is used to avoid numerical instabilities that arise from extreme parameter values. In addition, we generalize the dynamical system analysis formalism of Refs. \cite{2jeans_static,2jeans_cosmo,2jeans_short} to compute the Jeans scale for a system containing relativistic fluids, which essential for estimating the onset of gravitational instability. In the second part, we determine the critical curvature perturbation that sets the lower bound for PBH formation. To do so, we generalize the three‑zone model of Ref. \cite{Harda2013} to the two‑fluid case and analyze PBH formation using two different criteria, ultimately expressing the critical curvature perturbation as a function of the internal parameters of the three‑zone model. In the third part, we match the quantities in the quantum bouncing model at the time of gravitational instability to the internal parameters of the three‑zone model at the onset of collapse, and then compute the PBH mass fraction at formation.

\footnotetext{We refer to the Wentzel-Kramers-Brillouin calculation of the curvature power spectrum as “adiabatic” and that of the Jeans length as “WKB.” The term “adiabatic”  does not denote adiabatic perturbations; those are referred to here as curvature perturbations.}

This paper is organized as follows. Section~\ref{sec2} reviews the background evolution of dust–radiation bouncing cosmology and computes the curvature perturbation spectrum for the wavenumbers relevant for PBH formation. Section~\ref{sec3} derives the Jeans length for the two‑fluid model using a dynamical systems analysis. In Section~\ref{sec4} overdense regions are modeled as closed Robertson–Walker (cRW) patches and their dynamics are studied. Section~\ref{sec5} presents criteria for PBH formation in a cRW patch. Section~\ref{sec6} matches the quantities in the cRW model with those in the contracting universe and computes the PBH mass fraction for the relevant wavenumbers. Section~\ref{sec7} provides conclusions and discussion. Appendix~\ref{wkbaPP} derives adiabatic mode function for the curvature perturbation and the corresponding power spectrum. Appendix~\ref{B} presents the equations of motion for the density contrasts of the two fluids in the sub‑Hubble regime from the full perturbed Einstein equations and examines the validity of the WKB approximation used in the Jeans‑scale calculation.  Appendix~\ref{Appa} derives the sound‑wave front for the two‑fluid model.

Natural units $\hbar=c=1$ are used through out the paper.

\section{Dust-Radiation Bouncing Model}\label{sec2}

In this section we briefly review the dust–radiation quantum bouncing cosmology, covering the evolution of both the background and the perturbations. We then compute the curvature perturbation spectrum using the adiabatic method.

\subsection{Background Evolution}\label{Sec:backd}

In Ref. \cite{Barroso2025}, the  formation of PBH was investigated in a pure-dust quantum bouncing cosmological model. In the present study, we consider a more realistic setup with two fluids: a dust component characterized by a small equation of state parameter $w$, and  radiation with equation of state $1/3$.  Although PBH formation is analyzed in the classical contracting phase and does not depend on the details of the bounce, we outline the quantum bouncing mechanism for completeness.

We assume that near the singularity only one fluid dominates, which in this case is radiation. The quantum corrected Friedman equations near the quantum bounce were obtained using the quantum trajectory approach of Ref.~\cite{ACACIODEBARROS1998229} applied to wavefunctions that satisfy the Wheeler-De Witt equation. The effective Friedmann equation then reads
\begin{equation}
  \bar{\cal H}^{2} = \frac{8\pi G \bar{a}^2}{3}\left(\bar{\rho}_r - \frac{C_\textsc{q}}{\bar{a}^6}\right)\equiv \frac{8\pi G \bar{a}^2}{3}\left(\bar{\rho}_r - \rho_\textsc{q}\right),
  \label{QH2}
\end{equation}
where $C_{Q}$ is a constant, $\rho_Q$ is the effective quantum correction, and $\bar{\rho}_r$ is the energy density of radiation. Here $\bar{a}$ is the scale factor,  $\bar{\cal H}\equiv \bar{a}(\bar{\eta})'/\bar{a}(\bar{\eta})$, and the prime denotes a derivative with respect to the conformal time $\bar{\eta}$. 
Hereafter, unless stated otherwise, quantities with an overbar denote background quantities in the quantum bouncing model, distinguishing them from the corresponding quantities that appear in the three‑zone model discussed in Sects \ref{sec4} and \ref{sec5}. By adding the energy density of the dust fluid $\bar{\rho}_w$ in Eq. \eqref{QH2},  which does not affect the bounce, we obtain
\begin{equation}\label{Effective_Friedmann_2fluid}
    \bar{\cal H}^{2} = \frac{8\pi G \bar{a}^2}{3}\left(\bar{\rho}_{w}+ \bar{\rho}_{r} -
  \rho_\textsc{q} \right).
\end{equation}
The energy densities of dust, radiation and the quantum correction scale as $\bar{a}^{-3(1+w)}$, $\bar{a}^{-4}$, and $\bar{a}^{-6}$, respectively. Accordingly, Eq. \eqref{Effective_Friedmann_2fluid} can be rewritten as
\begin{equation}
    \bar{\cal H}^2 = \frac{8\pi G \bar{a}^2}{3}\dcha{ \bar{\rho}_w\dcol{1 - \dpar{\frac{\bar{a}_\textsc{b}}{\bar{a}}}^{3(1-w)} }+\bar{\rho}_r\dcol{1-\dpar{\frac{\bar{a}_\textsc{b}}{\bar{a}}}^2 } } \, ,\label{2fluid_Hubble}
\end{equation}
where $\bar{a}_\textsc{B}$ is the scale factor at the bounce point, defined by $\bar{\cal H}=0$. The energy densities $\bar{\rho}_w$ and $\bar{\rho}_r$ are normalized as
\bl
{\bar{\rho}_r=\bar{\rho}_{r,c}\left(\frac{\bar{a}_c}{\bar{a}}\right)^{4}},~~~
\bar{\rho}_w= \bar{\rho}_{w,c}\left(\frac{\bar{a}_c}{\bar{a}}\right)^{3(1+w)}.\label{sothat}
\el
The normalized scale factor in the contracting phase is $\bar{a}_c=0.625$,  relative to the current scale factor $\bar{a}_0=1$, which corresponds to the starting point of the numerical calculation in Ref. \cite{2fluid_pt1}. Quantities with the subscript $c$ are evaluated at the normalized time. 
Therefore, Eq.\eqref{2fluid_Hubble} can be rewritten as 
\bl
\Big[\frac{d}{dx}\left(\frac{\bar{a}}{\bar{a}_c}\right)\Big]^2 = \left(\frac{\bar{a}}{\bar{a}_c}\right)^4\dcha{\frac{\Omega_{w,c}}{(\bar{a}/\bar{a}_c)^{3(1+w)}}\dcol{1 - \dpar{\frac{\bar{a}_\textsc{b}/\bar{a}_c}{\bar{a}/\bar{a}_c}}^{3(1-w)} }+\frac{\Omega_{r,c}}{(\bar{a}/\bar{a}_c)^4}\dcol{1-\dpar{\frac{\bar{a}_\textsc{b}/\bar{a}_c}{\bar{a}/\bar{a}_c}}^2 } },\label{normalizetoknow}
\el
where the dimensionless rescaled conformal time $x\equiv (\bar{a}_c H_0 )\bar{\eta}$ is introduced. 
Here $H_0$ denotes the present Hubble rate, which takes the value $H_0\approx  69.560$ km s$^{-1}$ Mpc$^{-1}$ according to Ref.~\cite{PLANK}.
Note that the Hubble rate expressed in conformal time coincides with that in  cosmic time at present. The mass fractions of dust and radiation are defined 
as $\Omega_{w,c}\equiv \bar{\rho}_{w,c}/(3H_0^2/8\pi G)$ and $\Omega_{r,c}\equiv \bar{\rho}_{r,c}/(3H_0^2/8\pi G)$,
that is, the ratios of the dust and radiation  energy densities at the normalization point to the current critical energy density. In this paper we follow the representative fit to the Planck 2018 data (using baseline temperature and polarization angular power spectra:  $TT$-$TE$-$EE$-low-$\ell$) from Ref. \cite{2fluid_pt1}, and choose the  parameters in Table.\ref{paras}
\begin{table}[htp]
\centering
\begin{tabular}{|c|c|c|c|c|}
\hline
Parameters & $\Omega_{w,c}$ & $\Omega_{r,c}$ & $w$ & $\bar{a}_{\mathrm{B}}$ \\
\hline
Representative fits & $1-\Omega_{r,c}$ & $1.450\times 10^{-7}$ & $9.440\times 10^{-22}$ & $6.250\times 10^{-31}$ \\
\hline
\end{tabular}
\caption{Values of the parameters for the representative fits to the Planck 2018 data \cite{2fluid_pt1}.} \label{paras}
\end{table}

\begin{figure}[htb]     
\includegraphics[width = .5\linewidth]{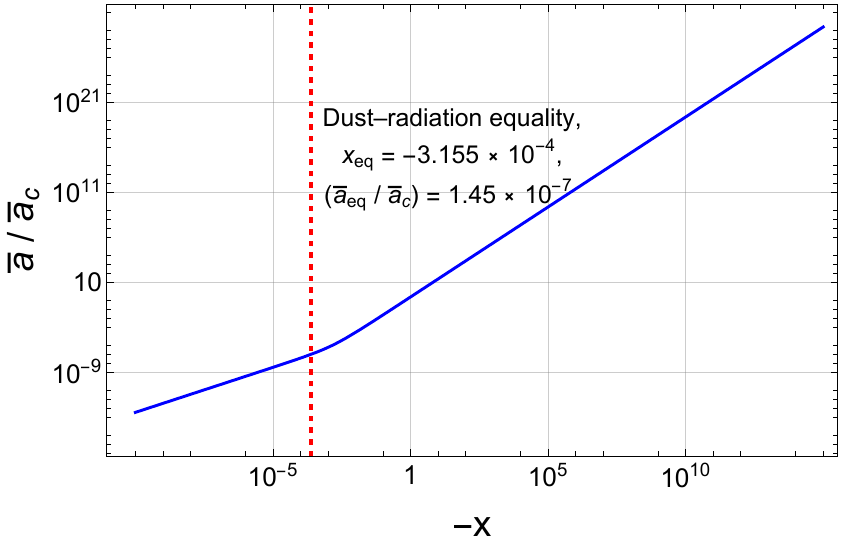}
 \centering
\caption{Blue: rescaled scale factor $\bar{a}/\bar{a}_c$ as a function of the dimensionless time $(-x)$. 
Red: dust-radiation equality. The evolution of the scale factor should be understood from the right hand side to the left in the contracting phase where $x<0$. The same applies to Figs.\ref{WKBvalid}, \ref{powerspectrumofiin}, and \ref{compare}.} \label{sf}
\end{figure}

With the boundary condition 
$\bar{a}(\bar{\eta}=0)=\bar{a}_\textsc{b}$, we numerically solve Eq.\eqref{normalizetoknow}. For illustration,
Fig.\ref{sf} shows the rescaled scale factor $(\bar{a}/\bar{a}_c)$ as a function of the dimensionless time $(-x)$.  
The red dashed line in Fig.\ref{sf} indicates the dust-radiation equality, which occurs when  $(\bar{a}_{\text{eq}}/\bar{a}_c)\approx (\Omega_{r,c}/\Omega_{w,c})\approx 1.45\times 10^{-7}$, corresponding to $x_{\text{eq}}=-3.155\times 10^{-4}$.

\subsection{Linear Perturbations}\label{sec22}

In this subsection, we derive the spectrum of the total curvature perturbation $\zeta$. This spectrum provides the variance of the Gaussian distribution assumed in the Press–Schechter formalism \cite{carr1974black,carr1975primordial}, and will be used to compute the PBH mass fraction in Sect.~\ref{sec6}.

By expanding the Einstein–Hilbert action coupled to the two fluids up to second order, without invoking the classical Einstein equation, one can derive a Hamiltonian that describes the dynamics of perturbations \cite{peter2007noninflationary,pinto2021bouncing}. In terms of the above variables, the Hamiltonian that describes scalar perturbations in Fourier space $\delta \mathcal{H}^{(2,s)} = \sum_k \delta
  \mathcal{H}^{(2,s)}_k$ for a given mode $k$ is given by Ref. \cite{Peter:2015zaa}
\begin{equation}
  \begin{split}
    \delta \mathcal{H}^{(2,s)}_k =  \frac{\Pi_{\zeta k}^2}{2m_\zeta}  +
    \frac{\Pi_{Q k}^2}{2m_Q}  + Y\Pi_{\zeta k}\Pi_{Q k}                +
    \frac{1}{2}m_{\zeta}\nu_{\zeta}^2  \zeta_k^2  +
    \frac{1}{2}m_Q\nu_Q^2 Q_k^2  \, ,
  \end{split}
  \label{2fluidhamiltonian}
\end{equation}
where $\zeta_k$ and $Q_k$ are the mode functions of the curvature and entropy perturbations, respectively, and $\Pi_{\zeta k}$ 
and $\Pi_{Q k}$ are the corresponding mode functions of the canonical momenta. They are defined as 
linear combinations of $\zeta_{wk}$ and $\zeta_{rk}$, the mode functions of the dust and
radiation curvature perturbations, together with  their first derivatives with respect to time (see
Refs. \cite{2fluid_pt1, Peter:2013avv, Peter:2015zaa} for their explicit definitions). The effective masses and frequencies of the curvature and entropy perturbations, $m_\zeta, m_Q,  \nu^{2}_{\zeta k}$, and $\nu^{2}_{Q k}$ are time dependent and given by 
\begin{subequations}
  \begin{align}
   m_\zeta & \equiv \frac{\bar{a}^4}{\bar{\cal H}^2} \frac{\left[ (1+w) \bar{\rho}_w+4/3 \bar{\rho}_r
        \right]^2}{w (1+w) \bar{\rho}_w+4/9 \bar{\rho}_r} \, ,\label{equaim1}
    \\
        m_Q& \equiv \frac{9\left[ (1+w)\bar{\rho}_w+4/3\bar{\rho}_r\right]^2}
    {4\bar{a}^4 (1+w)\bar{\rho}_w\bar{\rho}_r\left[ (1+w)\bar{\rho}_w + 4w \bar{\rho}_r\right]} \, ,  \label{equai0}
\\
    \nu^2_{\zeta k}& \equiv
    \frac{\left[w (1+w) \bar{\rho}_w+4/9 \bar{\rho}_r\right]}{(1+w) \bar{\rho}_w+4/3\bar{\rho}_r}
    k^2, \label{equai1}
    \\
    \nu^2_{Q k}  &\equiv \frac{1}{3}
    \frac{\left[(1+w) \bar{\rho}_w + 4 w \bar{\rho}_r\right]}{(1+w) \bar{\rho}_w +4/3\bar{\rho}_r}
    k^2, \label{equai2}
  \end{align}
\end{subequations}
and the effective coupling $Y$  defined as
\begin{equation}
  Y \equiv \frac{\bar{\cal H}(1+w)4/3(3w-1)\bar{\rho}_w\bar{\rho}_r}{3 \left[ (1+w)\bar{\rho}_w +
      (1+1/3) \bar{\rho}_r \right]^2}.
  \label{Ydef}
\end{equation}
By evaluating the Hamiltonian equations associated with  \eqref{2fluidhamiltonian} and eliminating the momenta, one obtains the equations of motion
\begin{subequations}\label{eqsofmotion}
\begin{align}
    \zeta_k'' + \bar{R}^{-1}_{\text{H}\zeta}\zeta_k' +
    \nu^2_ {\zeta k}\zeta_k & = - m_Q \nu^2_{Q k} Y Q_k +
    \frac{ (m’_\zeta / m_\zeta) Y
      + Y'}{1-m_\zeta m_Q Y^2}
    m_Q Q'_k,\label{10a}                       
    \\
    Q_k'' + \bar{R}^{-1}_{\text{H}Q}Q_k' + \nu^2_{Q k} Q_k & =
    - m_\zeta \nu^2_{\zeta k} Y \zeta_k +
    \frac{ (m_Q' / m_Q) Y + Y'}
    {1-m_\zeta m_Q Y^2 }
    m_\zeta \zeta'_k\, ,\label{10b}
  \end{align}
\end{subequations}
where the quantities $\bar{R}^{-1}_{\text{H}\zeta}$ and
$\bar{R}^{-1}_{\text{H}Q}$ are the effective Hubble radii  associated with the curvature 
and entropy perturbations, respectively, and are given by
\begin{equation}
     \bar{R}^{-1}_{H i}\equiv
    (m'_i/m_i
    + Y m_Q m_\zeta Y')( 1-m_\zeta m_Q Y^2 )^{-1}, \label{hubble_radiuses}
\end{equation}
where $i=(\zeta, Q)$. 
Note that, although dust and radiation are non‑interacting at the background level, see Eq. \eqref{2fluid_Hubble}. At the perturbative level, however, they become coupled through gravity, as shown in Eqs.\eqref{eqsofmotion}. To solve the coupled system one must specify initial conditions for both fluids. Because of the interaction between them, the standard adiabatic initial conditions for free fields are no longer appropriate. A consistent choice can instead be defined using the coupled adiabatic vacuum prescription proposed in Ref.~\cite{Peter:2015zaa} and applied to the dust–radiation system in Ref.~\cite{2fluid_pt1}. This construction generalizes the usual adiabatic vacuum to the coupled case, providing a physically motivated set of initial conditions for Eqs.~\eqref{eqsofmotion}, from which the curvature perturbation spectrum  ${\cal P}_{\zeta}(k)$ can be obtained.

As shown in Ref. \cite{2fluid_pt1}, however, the entropy  spectrum ${\cal P}_{Q}(k)$ is much smaller than the curvature one. The mechanism is dynamical and can be understood as follows. In a single fluid dominated stage in the contracting phase, the effective masses scale as $m_{\zeta} \propto \bar{a}^{2}, m_{Q} \propto \bar{a}^{-4}$. Thus, the mass of the curvature mode $m_{\zeta}$ decreases continuously in the contracting phase, whereas the mass of the entropy mode $m_{Q}$ increases. By analogy with a harmonic oscillator, the amplitude grows when the  mass decreases and freezes when the mass increases. Consequently, the amplitude of the curvature mode $\zeta$ grows during contraction, while the entropy mode $Q$ remains frozen. The curvature mode therefore naturally emerges with a much larger amplitude than the entropy mode, without any fine‑tuning.

Since the curvature mode is dominant in the contracting phase, in order to conveniently  solve Eqs.\eqref{eqsofmotion} under the extremely small parameters given in Table \ref{paras}, we neglect the entropy modes by setting $Q_{k} =0 $ and $ Y = 0$. Under this approximation, Eq. \eqref{10a} reduces to
\begin{equation}
    \zeta''_k + \frac{m_{\zeta}'}{m_{\zeta}}\zeta'_k +
    \nu^2_\zeta \zeta_k  = 0,  \,   \label{eq14}            
\end{equation}
while Eq. \eqref{10b} becomes an identity, $0=0$. Applying the change of variable
\bl
\zeta_k=\frac{u_k}{\sqrt{m_{\zeta}}},\label{djew}
\el
to Eq. \eqref{eq14} yields 
\bl
\ddot{u}_{k}+\left[\nu_{\zeta n}^2-\frac12\frac{\ddot{m}_{\zeta n}}{m_{\zeta n}}+
\frac14 \left(\frac{\dot{m}_{\zeta n}}{m_{\zeta n}}\right)^2\right]u_{k}=0,\label{ULH}
\el
where dots denotes the derivative $d/dx$. To simplify the calculation, we have introduced 
the following 
normalized effective frequency $\nu_{\zeta n}$ and mass $m_{\zeta n}$
\bl
\nu_{\zeta n}\equiv \frac{\nu_{\zeta}}{\bar{a}_c H_0}
=\left[1+\frac{4(1/(3w)-1)\Omega_{r,c}(\bar{a}_c/\bar{a})^4}{3(1+w)\Omega_{w,c}(\bar{a}_c/\bar{a})^{3(1+w)}+4\Omega_{r,c}(\bar{a}_c/\bar{a})^4}\right]^{1/2} \sqrt{w} k_H,\label{effecitfreqn}
\el
and 
\bl
m_{\zeta n} \equiv \frac{m_{\zeta } G}{\bar{a}_c^{2}}  
=\frac{3  }{8\pi }\left(\frac{\bar{a}}{\bar{a}_c}\right)^2\frac{\left( (1+w) 
\Omega_{w,c}(\bar{a}_c/\bar{a})^{3(1+w)}+\frac43 \Omega_{r,c}(\bar{a}_c/\bar{a})^4
        \right)^2}{\left(\Omega_{w,c}(\bar{a}_c/\bar{a})^{3(1+w)}+ \Omega_{r,c}(\bar{a}_c/\bar{a})^4\right)\left( w (1+w) 
\Omega_{w,c}(\bar{a}_c/\bar{a})^{3(1+w)}+\frac49 \Omega_{r,c}(\bar{a}_c/\bar{a})^4
\right)}.\label{mzetatfreqn}
\el
Here $k_H\equiv k/(\bar{a}_c H_0)$ is the dimensionless rescaled wavenumber. Equations
\eqref{equaim1} and \eqref{equai1}, together with 
the definitions of $\Omega_{w,c}$ and $\Omega_{r,c}$, were used.
To numerically solve Eq. \eqref{mzetatfreqn}, 
we impose the following initial mode function,
\bl
\lim_{\bar{\eta}\rightarrow-\infty} u_k
&= \frac{ 1}{\sqrt{4 \sqrt{w} (\bar{a}_c H_0) k_H}}e^{-i \sqrt{w}k_H x},\label{dhewuiuh28}
\el
which ensures that the vacuum energy density coincides with that of a harmonic 
oscillator for each mode, $\rho_k\equiv |u_k'|^2+k^2|u_k|^2=\frac{\sqrt{w}k}{2}$. This condition is required because the curvature perturbation originates from quantum vacuum fluctuations, and the energy of each 
$k$-mode must match that of a harmonic oscillator.

In what follows, we derive the 
spectrum of the curvature perturbation 
for four rescaled wavenumbers
shown in Table. \ref{tablm1}, they 
correspond to various PBH masses.
PBHs in the large‑mass range can act as seeds of supermassive black holes~\cite{Carr2025}. PBHs of intermediate mass are candidate sources for gravitational‑wave events and may populate the mass gap in the stellar‑mass black‑hole spectrum~\cite{Andres-Carcasona:2024wqk}. In the light‑mass window, PBHs may constitute the entirety of dark matter~\cite{Katz:2018zrn, Carr:2020gox}. For the extremely small‑mass range, formation occurs around the dust-radiation equality.

\begin{table}[htbp] 
\centering \begin{tabular}{|c|c|} 
\hline 
Rescaled wavenumber & PBH mass
\\ \hline 5.372 $\times$ 10$^{3}$ & 10$^{13}M_{\odot}$ 
\\ \hline 2.493 $\times$ 10$^{7}$ & 10$^{2}M_{\odot}$
\\ \hline 5.377 $\times$ 10$^{12}$ & 10$^{-14}M_{\odot}$
\\ \hline 5.623 $\times$ 10$^{13}$ & 10$^{-17}M_{\odot}$
\\ \hline
\end{tabular} 
\caption{Correspondence between rescaled wavenumbers and PBH masses.} \label{tablm1}
\end{table}

\begin{figure}[htb]   
\includegraphics[width = .65\linewidth]{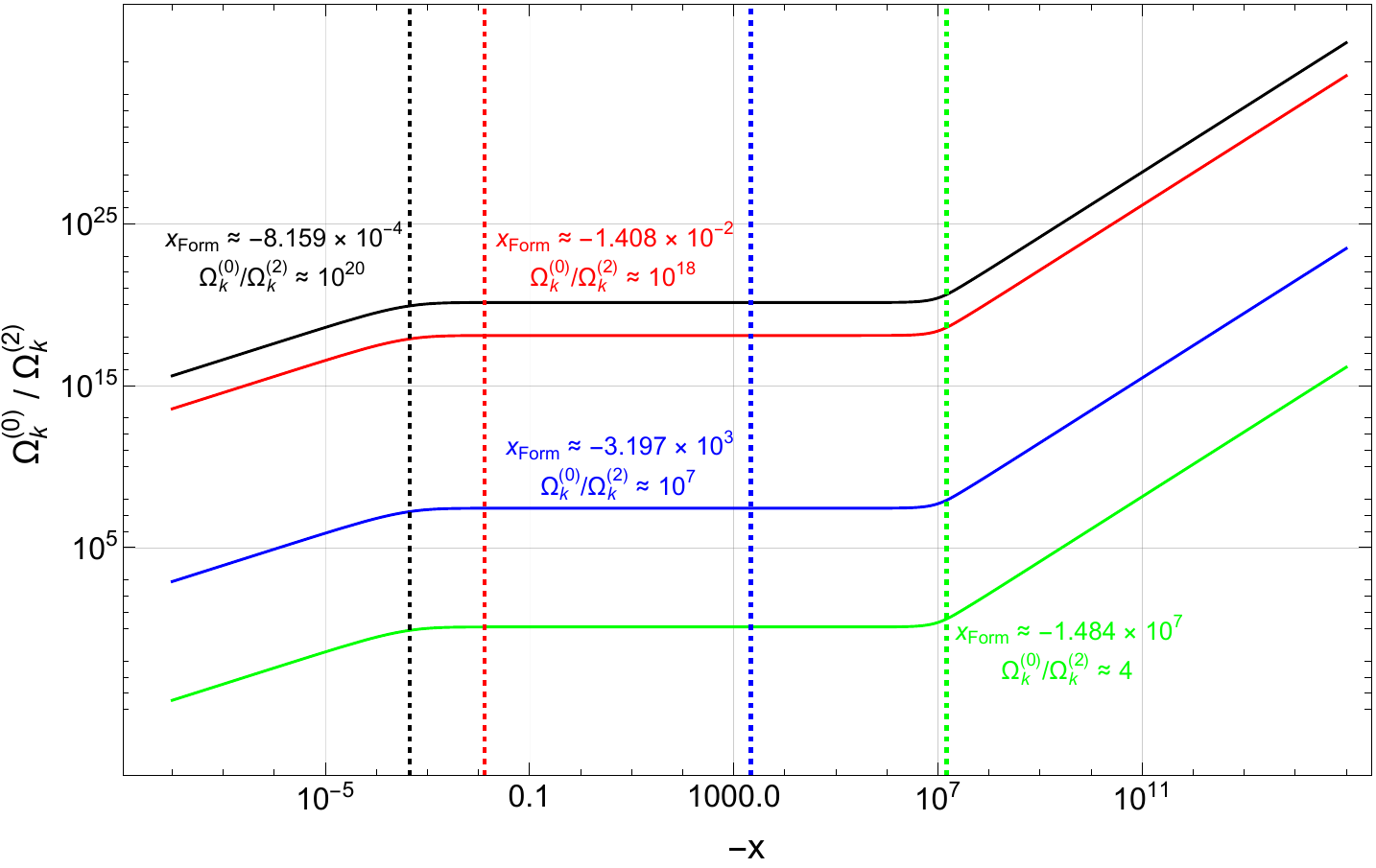}
 \centering
\caption{$\Omega_k^{(0)}/\Omega_k^{(2)}$ as a function of the dimensionless time. Green: PBH with 
mass $10^{13} M_{\odot}$. Blue: PBH with 
mass $10^{2} M_{\odot}$. Red: PBH with 
mass $10^{-14} M_{\odot}$. Black: PBH with 
mass $10^{-17} M_{\odot}$. The value of $\Omega_k^{(0)}/\Omega_k^{(2)}$
at the formation time is indicated by the vertical dashed lines.} \label{WKBvalid}
\end{figure}

Eq.~\eqref{ULH} can be solved using the adiabatic approximation order by order, with details given in Appendix~\ref{wkbaPP}. The adiabatic mode function up to second order is~\cite{Chakraborty,Parker_Toms_2009,Zhang2020}
\bl
u_{k} (x)\approx\frac{ 1}{\sqrt{4\bar{a}_cH_0 \left(\Omega_{k}^{(0)}(x)
+\Omega_{k}^{(2)}(x)
\right)}
} e^{-i \int^{x}\left(\Omega_{k}^{(0)}(x')
+\Omega_{k}^{(2)}(x')
\right) dx'},\label{modefuntionapproxiamtion}
\el
where the 0th order and 2nd order effective frequencies, $\Omega_{k}^{(0)}$
and $\Omega_{k}^{(2)}$, are given by Eqs.\eqref{0th} and \eqref{Omnewgkh2}, respectively. 
The adiabatic method requires $\Omega_{k}^{(0)}\gg\Omega_{k}^{(2)}$. To demonstrate 
the validity of this approximation,  in Fig. \ref{WKBvalid}, 
we plot $\Omega_k^{(0)}/\Omega_k^{(2)}$ for the 
rescaled wavenumbers defined in Table. \ref{tablm1}.
For the masses $10^{2}, 10^{-14}$ and $ 10^{-17} M_{\odot}$, the adiabatic condition is well satisfied, $\Omega_k^{(0)}/\Omega_k^{(2)}\gg1$ (see the black, red and blue thick curves in Fig.\ref{WKBvalid}), even after the PBH formation time (indicated by the vertical dashed lines in Fig. \ref{WKBvalid}). The PBH formation times are computed in Sect.\ref{sec6}. For the super massive case $10^{13} M_{\odot}$, the adiabatic approximation becomes 
marginal at its formation time (green vertical dashed line) but remains usable with $\Omega_k^{(0)}/\Omega_k^{(2)}\approx 4$. We therefore expect the exact solution to deviate only modestly from the adiabatic result. 

The dimensionless curvature perturbation spectrum is defined as 
\bl
{\cal P}_{\zeta }(k,\bar{\eta})&\equiv
\frac{k^3}{2\pi^2}|\zeta_k|^2
= (G H_0^2 ) (H_0 \bar{a}_c)   \frac{k_H^3  }{2\pi^2 m_{\zeta n }}|u_k|^2,\label{pzeta}
\el
where the right hand side is also dimensionless. In natural units,
the combination $(G H_0^2)$ is dimensionless, $(H_0 \bar{a}_c)$ has the dimension of inverse time, and $|u_k|^2$ 
has the dimension of time, as follows from the initial mode function in Eq.~\eqref{dhewuiuh28} or the adiabatic mode function in Eq.~\eqref{modefuntionapproxiamtion}. Substituting \eqref{modefuntionapproxiamtion} into \eqref{pzeta} and keeping the 
2nd order terms  yields 
\bl
{\cal P}_{\zeta}(k_H,x)
&\approx \frac{ (G H_0^2)k_H^3}{8\pi^2 m_{\zeta n} \nu_{\zeta n}}\left(1+\frac1{4\nu^2_{\zeta n}}\frac{\ddot{m}_{\zeta n}}{m_{\zeta n}}-
\frac1{8\nu^2_{\zeta n}} \left(\frac{\dot{m}_{\zeta n}}{m_{\zeta n}}\right)^2+
\frac{ \ddot{\nu}_{\zeta n}}{4\nu_{\zeta n}^3}
-\frac{3 \dot{\nu}_{\zeta n}^2}{8\nu_{\zeta n}^4}\right),\label{dewhiutext}
\el
where Eqs. \eqref{0th} and \eqref{Omnewgkh2} have been used.
\begin{figure}[htb]
 \subcaptionbox{}
   {%
     \includegraphics[width = .48\linewidth]{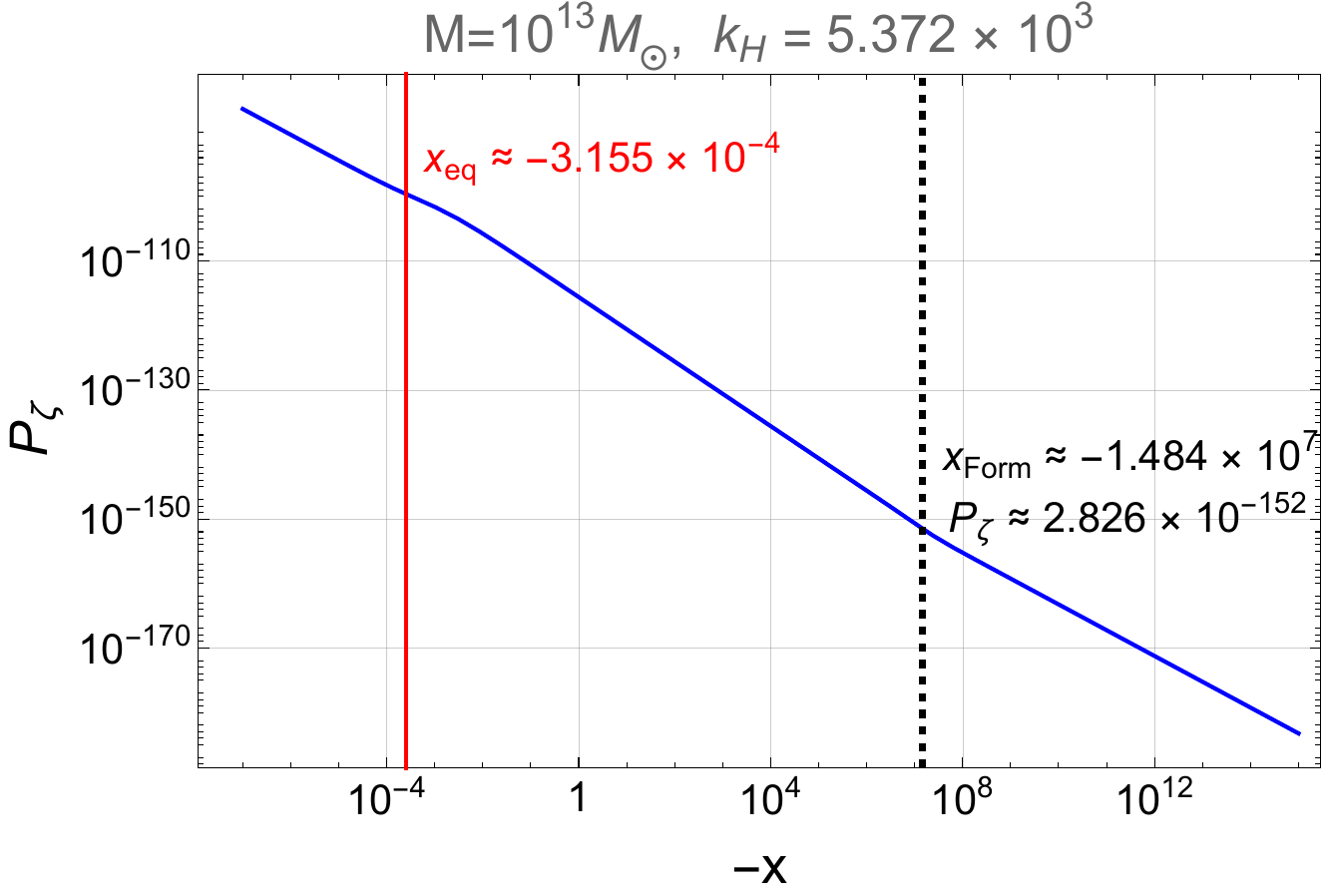}}
 \centering
 \subcaptionbox{}
   {%
     \includegraphics[width = .48\linewidth]{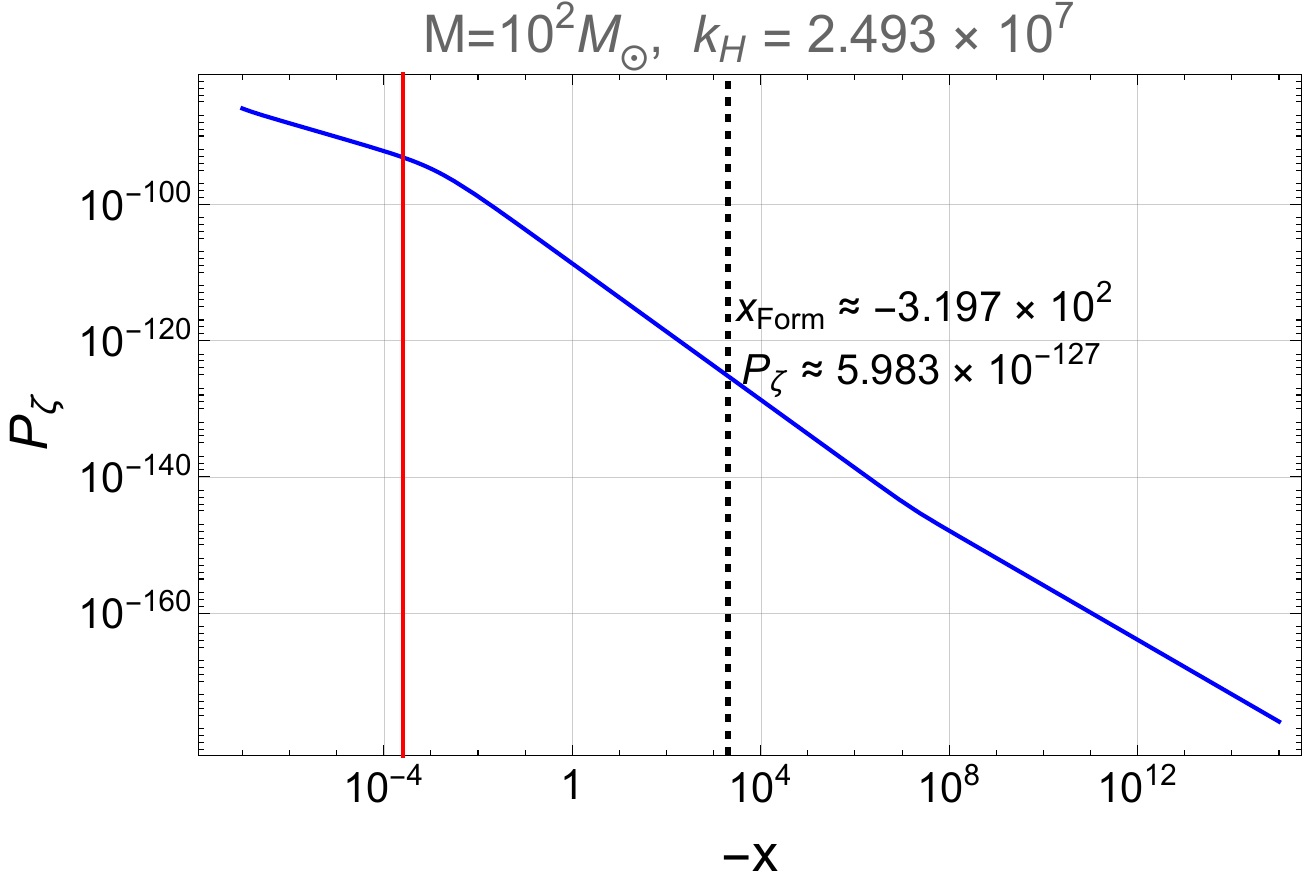}}
 \subcaptionbox{}
   {%
     \includegraphics[width = .48\linewidth]{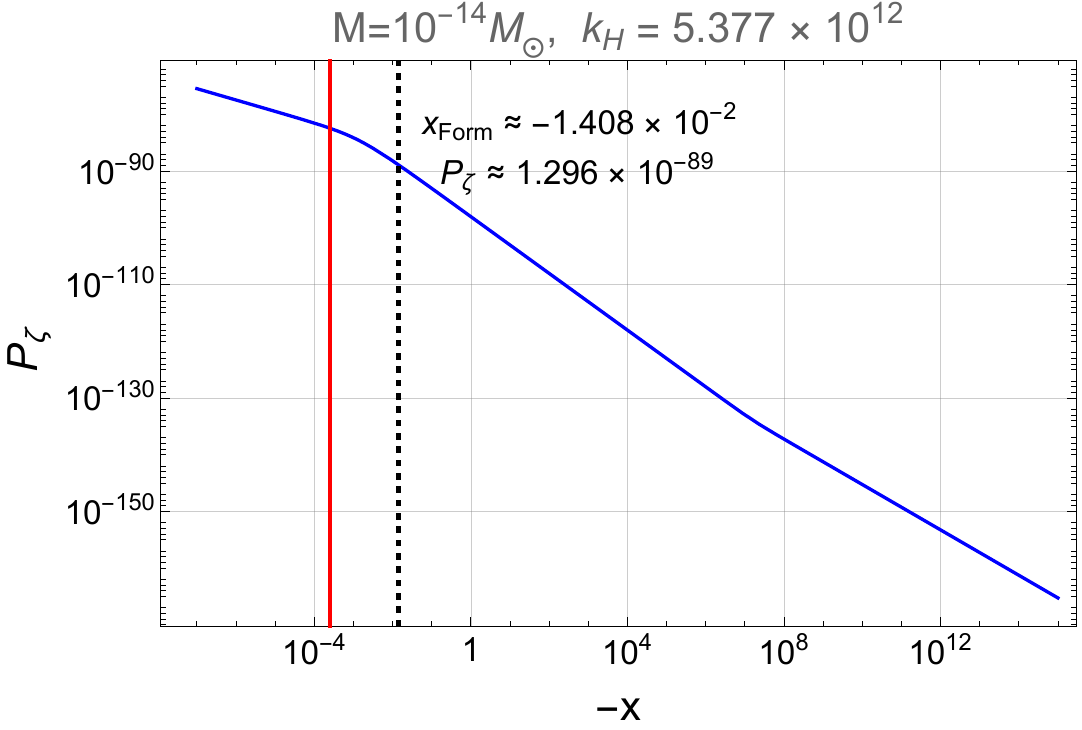}}
     \subcaptionbox{}
   {%
     \includegraphics[width = .48\linewidth]{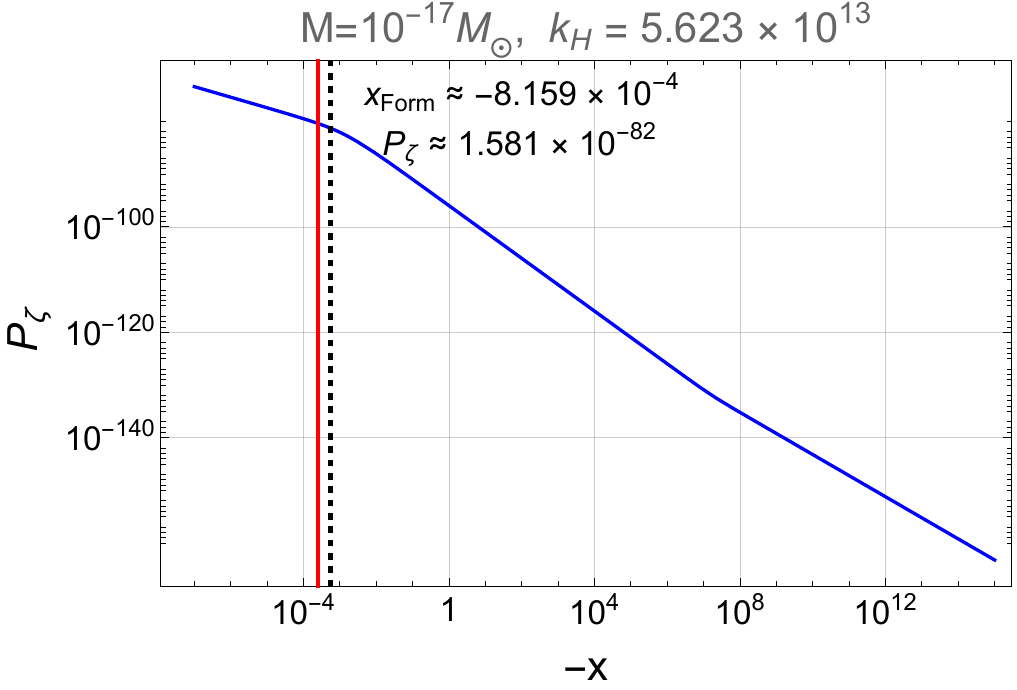}}
\caption{ Blue: Evolution of spectrum ${\cal P}_{\zeta}(k_H,x)$ at a given scale. Red: dust-radiation equality. Black: PBH formation time. (a)  $M=10^{-14} M_{\ensuremath{\odot}}$. (b)  $M=10^{2} M_{\ensuremath{\odot}}$  (c) $M=10^{13} M_{\ensuremath{\odot}}$. 
(d) $M=10^{-17} M_{\ensuremath{\odot}}$. } \label{powerspectrumofiin}
\end{figure}
In Eq.\eqref{dewhiutext}, the quantities $G H_0^2$, $k_H$, $m_{\zeta n}$, $\nu_{\zeta n}$ and $(d/dx)$ are dimensionless in natural units, and hence the spectrum of the curvature perturbation ${\cal P}_{\zeta}$ derived using the WKB approximation is 
dimensionless as well.  

The curvature power spectra for the rescaled wavenumbers defined in Table.\ref{tablm1} are shown in Figs.~\ref{powerspectrumofiin}(a)–(d) (blue curves). The red vertical lines in Figs.~\ref{powerspectrumofiin} mark dust–radiation equality. The spectrum  grows during the contracting phase. PBHs of smaller mass form at later times, indicated by the black vertical lines, and therefore exhibit a larger power spectrum at formation. 

We have examined the background evolution and the 
evolution of the $k$-mode curvature perturbations during the contracting phase of the two‑fluid quantum bouncing model. To determine which perturbation scales can undergo collapse at a given time, the next section analyzes the Jeans stability of the two‑fluid system.

\section{Dynamical-systems Analysis of the Jeans Length}\label{sec3}

The Jeans length, derived from linear perturbation theory, characterizes gravitational instability: density contrast modes with wavelengths larger than the Jeans length grow exponentially due to gravitational collapse. This scale links the quantum bouncing model discussed in Sect. \ref{sec2} with the three-zone model introduced in Sect. \ref{sec4}. In this section, we derive the Jeans scale for the two-fluid system.

The standard method to obtain the Jeans length for a single fluid (or when the second fluid is included to provide an additional gravitational potential) is to set the coefficient of the density contrast in its equation of motion to zero~\cite{Jones2017}. This approach is invalid when the dust–radiation system is treated as a coupled whole. We therefore employ the dynamical-system analysis of Refs.~\cite{2jeans_static,2jeans_cosmo,2jeans_short} to study the Jeans length in a two‑fluid system. Unlike Refs.~\cite{2jeans_static,2jeans_cosmo,2jeans_short}, where the density‑contrast equations are introduced by assumption, our treatment follows directly from the perturbed Einstein equations (see Appendix~\ref{B} for details). We begin with the coupled equations valid in the sub‑Hubble regime
\begin{subequations}\label{Dsys}
    \begin{align}
      \tilde{\delta}_w^{\prime\prime}+\left[ w k^2{+\frac{2}{3}\pi G\bar{a}^2(\bar{\rho}_r-6\bar{\rho}_w)}\right]\tilde{\delta}_w &= 8\pi G \bar{a}^{5/2}\bar{\rho}_{r}\tilde{\delta}_{r} \, , \label{app_dynamical_system1}\\ 
     \tilde{\delta}_r^{\prime\prime}+\frac{1}{3}\left( k^2 -{32}\pi G \bar{a}^2 \bar{\rho}_{r}\right)\tilde{\delta}_r &= {\frac{16}{3}}\pi G \bar{a}^{3/2}\bar{\rho}_{w}\tilde{\delta}_{w} \, , \label{app_dynamical_system2}
    \end{align}
\end{subequations}
where $\tilde{\delta}_{w,k}\equiv \bar{a}^{\frac{1-3w}{2}}\delta_w$ and 
$\tilde{\delta}_{r,k}\equiv\delta_{r,k}$ denote the rescaled 
density contrast modes for dust and radiation, respectively. 
The coupled equations in \eqref{Dsys} can be rewritten in matrix form
as
\begin{equation}
\label{diagonalize_system}
    \textbf{x}^\prime = \textbf{T} \textbf{x},
\end{equation}
where the column vector \textbf{x} is defined by $
    \textbf{x} \equiv (\tilde{\delta}^{\prime}_{w},
    \tilde{\delta}^{}_{w},
    \tilde{\delta}^{\prime}_{r},
    \tilde{\delta}_{r})^{\text{T}}$
and the matrix \textbf{T} is
\begin{equation}\label{MartixT}
\textbf{T} =     
\begin{pmatrix}
   0 & -\left[ w k^2+\frac{2}{3}\pi G\bar{a}^2(\bar{\rho}_r{-6\bar{\rho}_w})\right] & 0 & 8\pi G \bar{a}^{5/2}\bar{\rho}_{r} \\
    1 & 0 & 0 & 0 \\
    0 & {\frac{16}{3}\pi G \bar{a}^{3/2}\bar{\rho}_{w}} & 0 & -\frac13(  k^2 -32\pi G \bar{a}^2 \bar{\rho}_{r}) \\
    0 & 0 & 1 & 0
  \end{pmatrix}
.\end{equation}
Introducing the vector $\textbf{y}=\textbf{A}^{-1} \textbf{x}$, Eq. \eqref{diagonalize_system} becomes
\begin{equation}
    \textbf{y}^\prime = \textbf{A}^{-1}\textbf{T}\textbf{A}\textbf{y} - \textbf{A}^{-1}\textbf{A}^{\prime}\textbf{y} \, ,\label{fjheurih}
\end{equation}
where $\textbf{A}^{-1}\textbf{T}\textbf{A}$ is diagonalized 
and $\textbf{A}$ is constructed from the eigenvectors $\bm{\xi}_j$ of
$\textbf{T}$, i.e. $\textbf{A}\equiv (\bm{ \xi}_1,\bm{ \xi}_2,\bm{ \xi}_3,\bm{ \xi}_4)$. 
The explicit expressions for $\bm{ \xi}_j$ are given in Appendix \ref{subb2}. The diagonal entries of
$\textbf{A}^{-1}\textbf{T}\textbf{A}$  are 
\begin{subequations}\label{eigv}
    \begin{align}
        \lambda_1 = - \lambda_2& = -\frac{1}{\sqrt{2}}\sqrt{f(k,\bar{a}) - \sqrt{f(k,\bar{a})^2+4g(k,\bar{a})}}\label{eigv1t} \, ,
        \\
        \lambda_3 = -\lambda_4& = -\frac{1}{\sqrt{2}}\sqrt{f(k,\bar{a}) + \sqrt{f(k,\bar{a})^2+4g(k,\bar{a})}}\label{eigv3t}\, ,
    \end{align}
\end{subequations}
with the functions $f(k, \bar{a})$ and $ g(k,\bar{a})$ defined as
\begin{subequations}
    \begin{align}
        f(k,\bar{a}) & \equiv
        2\pi G\bar{a}^2(2\bar{\rho}_w+5\bar{\rho}_r)-\dpar{w+1/3}k^2 \, , \\ 
        g(k,\bar{a})&\equiv\frac{1}{3} \left\{128 \pi ^2 G^2 \bar{a}^4 \bar{\rho}_w \bar{\rho}_r-\big(k^2-32 \pi  G \bar{a}^2 \bar{\rho}_r\big) \left[k^2 w+\frac{2}{3} \pi  G \bar{a}^2 ( \bar{\rho}_r-6\bar{\rho}_w )\right]\right\} \, . 
    \end{align}
\end{subequations}
When the matrices $\textbf{A}^{-1}\textbf{T}\textbf{A}$ and $\textbf{A}$ are time independent, the solution of Eq. \eqref{fjheurih} is given by \cite{2jeans_static,2jeans_cosmo,2jeans_short}
\bl
\boldsymbol{y}=\sum_{j=1}^4 \alpha_j \mathrm{e}^{\lambda_j \bar{\eta}} \boldsymbol{\xi}_j,\label{capitaly}
\el
where the coefficients $\alpha_j$ are determined by initial conditions. In the present case, $\textbf{A}^{-1}\textbf{T}\textbf{A}$ and $\textbf{A}$ are time dependent. Nevertheless, under the WKB approximation the leading-order solution for $\textbf{y}$ retains the same functional form, with the eigenvalues $\lambda_i$ given by Eq. \eqref{eigv}. Higher-order corrections within the WKB expansion are presented in Appendix \ref{subb2}, where the validity of the approximation is also examined.
\begin{figure}[htb]
   {%
     \includegraphics[width = .7\linewidth]{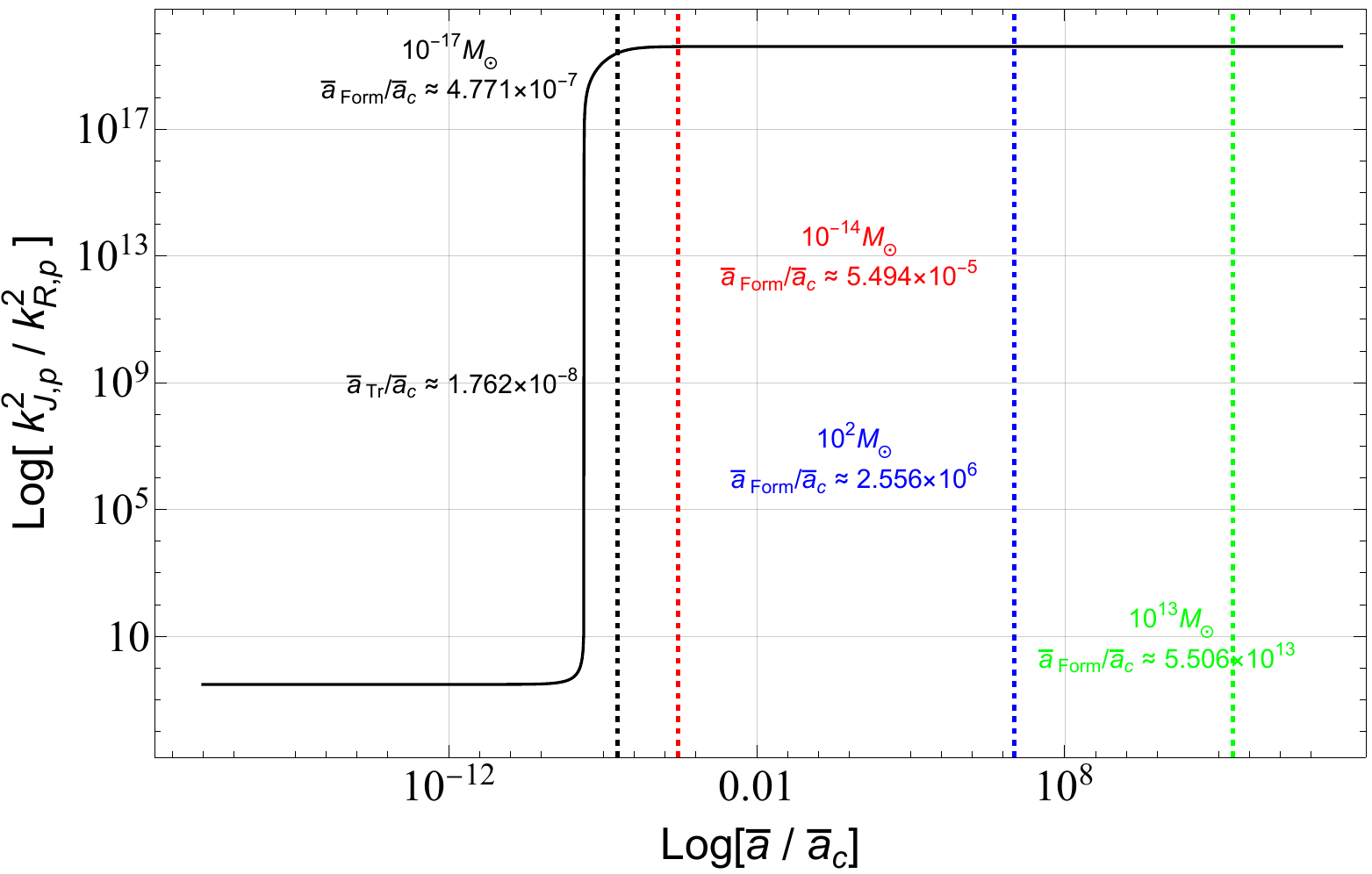}}
 \centering
\caption{Comparison between the Jeans wavenumber and the Hubble wavenumber.} \label{compare}
\end{figure}
The wavenumber satisfying $\lambda_i=0$ is the Jeans wavenumber, 
which defines the transition between oscillatory and growing modes.
Although $\lambda_i=0$ yields two roots, we consider only the physically relevant branch
\begin{equation}
    k_{J}^2=\frac{\pi  G \bar{a}^2}{3 w}\left[ 6\bar{\rho}_w-(1-48 w) \bar{\rho}_r +\sqrt{( 6\bar{\rho}_w-(1-48 w) \bar{\rho}_r)^2+192 w \bar{\rho}_r^2}\right],\label{jeanss2t}
\end{equation}
which is positive in the parameter range of interest. The wavenumber \eqref{jeanss2t} reduces to the single-component results for dust and radiation when
$\bar{\rho}_r$ and $\bar{\rho}_w$ are set to zero, respectively,
\begin{subequations}
\bl
\lim_{\bar{\rho}_r\rightarrow0} k_{J}^2&=4\pi G \bar{a}^2w^{-1} \bar{\rho}_w,\label{limitrhow0}
\\
\lim_{\bar{\rho}_w\rightarrow0} k_{J}^2&=
32\pi G  \bar{a}^2 \bar{\rho}_r.\label{limitrhor0}
\el
\end{subequations}
Using Eq. \eqref{2fluid_Hubble}, together with $\Omega_{w,c}\equiv \bar{\rho}_{w,c}/(3H_0^2/8\pi G)$ and $\Omega_{r,c}\equiv \bar{\rho}_{r,c}/(3H_0^2/8\pi G)$, 
the Jeans wavenumber can be rewritten as 
\bl
k_J=\frac{\bar{a}H_0}{\sqrt{8w}} \left[\frac{6 \Omega_{w, c}}{(\bar{a}/\bar{a}_c)^{3 (1+w)}}-\frac{(1-48 w) \Omega_{r, c}}{(\bar{a}/\bar{a}_c)^4}+\sqrt{\left(\frac{6 \Omega_{w, c}}{(\bar{a}/\bar{a}_c)^{3 (1+w)}}-\frac{(1-48 w) \Omega_{r, c}}{(\bar{a}/\bar{a}_c)^4}\right)^2+\frac{192 w \Omega_{r,c}^2}{(\bar{a}/\bar{a}_c)^8}}\right]^{1/2}.\label{kJcomi}
\el
We now assess the regime of validity of this Jeans wavenumber. In deriving Eq.~\eqref{Dsys}, we have assumed that the modes are sub‑Hubble, i.e. the comoving wavenumber satisfies $k \gg\bar{\cal H}$. In terms of physical wavenumbers, this corresponding to
$k_{p}\equiv k/\bar{a}\gg\bar{\cal H}/\bar{a}$. To compare the physical Jeans wavenumber $k_{J,p}=k_J/\bar{a}$ with 
the physical Hubble wavenumber, we define 
\bl
 k_{R,p}\equiv\frac{2\pi}{(\bar{a}/\bar{{\cal H}})}=
2\pi H_0\sqrt{\Omega_{w,c}(\bar{a}_c/\bar{a})^{3(1+w)}+\Omega_{r, c}(\bar{a}_c/\bar{a})^4},
\el
where Eq.\eqref{2fluid_Hubble} has been used and quantum corrections are neglected. Fig. \ref{compare} shows the ratio 
$k_{J,p}^2 / k_{R,p}^2$. After the transition scale factor
$\bar{a} < \bar{a}_\text{Tr}$, the Jeans wavenumber fails below the Hubble wavenumber. Perturbations thus become super-Hubble before becoming super-Jeans.
Therefore, this work considers 
PBH formation in the regime $\bar{a} > \bar{a}_\text{Tr}$. The  PBHs of interest in this paper lie before the transition scale factor, as indicated by the four vertical lines in Fig. \ref{compare}. The transition scale factor is  obtained by solving  $k_{J, p}=k_{R, p}$. The physically relevant solution is   
\bl
\bar{a}_\text{Tr}/\bar{a}_c\approx \frac{\Omega_{r,c}}{\Omega_{w,c}}\frac{  (48-32 \pi ^2) w+5   -\sqrt{ 96 w (24 w-5)+49-72/\pi^2}}{4 (8 \pi ^2 w-3)}\approx  1.762 \times 10^{-8},\label{atr}
\el
where the term $(\bar{a}/\bar{a}_c)^w$ has been neglected for simplification. The equation  $k_{J, p}=k_{R, p}$ also yields another root with  $\bar{a}<0$, which is unphysical and therefore discarded. The transition scale factor $\bar{a}_\text{Tr}/\bar{a}_c$ is close to radiation–matter equality, $(\bar{a}_\text{eq}/\bar{a}_c)\approx (\Omega_{r,c}/\Omega_{w,c})\approx 1.45\times 10^{-7}$, differing by an $O(10^{-1})$ factor.   

We have
computed the amplitudes of the curvature‑perturbation modes on the relevant scales, and determined which perturbation scales become gravitationally unstable and begin to collapse at a given time. In the following section we employ the three‑zone model to describe the subsequent matter collapse.

\section{Three-Zone Model: Density Contrast Evolution}\label{sec4}

In this section, to describe the evolution of the density contrast in a system containing both dust and radiation, we generalize the three-zone model proposed by Harada et al. \cite{Harda2013} from the single- to the two-fluid case. This generalized model will be later matched with 
the quantum bouncing model in Sect.\ref{sec6}.

The three-zone model consists of three regions. The outermost region is described by the flat Robertson–Walker (fRW) spacetime,
\bl
ds^2=\bar{a}(\bar{\eta})^2(-d\bar{\eta}^2+ r^2 d\Omega^2),\label{outside}
\el
where $r$ is 
the radial coordinate, $d\Omega$ denotes the metric on the unit two-sphere, and $\bar{a}(\bar{\eta})$ is the background 
scale factor obtained from the solution of Eq.\eqref{2fluid_Hubble} and shown in Fig. \ref{sf}. The innermost overdense region is described by the 
closed Robertson-Walker (cRW) spacetime   \cite{Harda2013,Kopp2011}
\bl
ds^2=A(\eta')^2(-d\eta'^2+ d\chi^2+\sin\chi^2 d\Omega^2),\label{inter}
\el
where $\eta'$ is the conformal time of the overdense region, $A(\eta')$ is the scale factor (with  dimension of length), and $\chi$ is related to the radial coordinate 
\bl
r'= \sin\chi.\label{cdhweuih}
\el
Let $\chi_a$ denote the boundary coordinate of the overdense region. It may take values in either of the ranges 
\bl
\chi_a\in[0,\pi/2] \text{$~$or$~$} \chi_a\in[\pi/2,\pi].\label{eq37}
\el
For the following discussion, we choose the former case, referred to as Type I in Ref.\cite{Kopp2011}. Plugging Eq. \eqref{cdhweuih} into Eq. \eqref{inter}  yields  \cite{Harda2013,Kopp2011}
\bl
ds^2
&=A(\eta')^2\left(-d\eta'^2+ \frac{dr'^2}{1-r'^2}+r'^2 d\Omega^2\right),\label{interorigin}
\el
which is the standard the cRW metric. 
To convert the dimensional scale factor $A(\eta')$ to the dimensionless one, 
we introduce a characteristic length scale $L$ and define the dimensionless variables (following Ref. \cite{Carroll2019}):
\begin{subequations}
\bl
A(\eta') &\rightarrow a(\eta')=A(\eta')/L,\label{repla1}
\\
r'&\rightarrow r= L r',
\\
\eta' &\rightarrow \eta=L\eta'.
\el    
\end{subequations}
Under these rescalings, the metric \eqref{interorigin} 
becomes 
\bl
ds^2&=a(\eta)^2\left[-d\eta^2+ \frac{dr^2}{1-(r/L)^2}+r^2 d\Omega^2\right],\label{eqwhuiwrfh}
\el
where $L$ is an undetermined constant 
with dimensions of length. Its value will be fixed when the three-zone model is matched to the quantum bouncing model described in Sect. \ref{sec2}.  The Friedman equation for the cRW region follows from Eq.\eqref{eqwhuiwrfh} and is given by 
\bl
\left(\frac{a'}{a}\right)^2=\frac{8\pi G\rho_\text{eq}}{3}\left(\frac{a_\text{eq}^{3(1+w)}}{a^{3(1+w)}}+\frac{a_\text{eq}^4}{a^4}\right) a^2-\frac{1}{L^2}, \label{closlefreihidjewoi}
\el
where $\rho_\text{eq}$ is the energy density of the cRW at dust-radiation equality, and $a_\text{eq}$ is the dimensionless scale factor at that epoch. The dust equation of state $w$ is $ 9.44\times10^{-22}$, which is the same as background quantity in Table \ref{paras}.

\begin{figure}[htb]
 \centering
   {%
     \includegraphics[width = .5\linewidth]{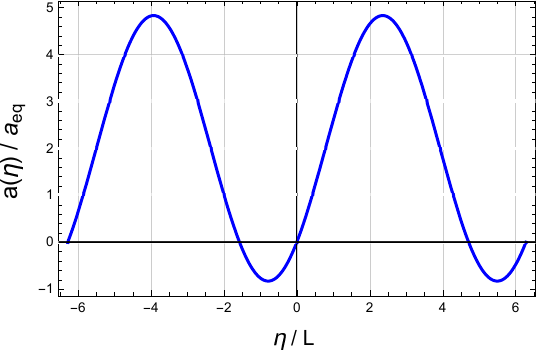}}
\caption{ Scale factor of the overdense region for
$\eta_\text{eq}/ L=1$.} \label{intermorefer}
\end{figure}

To simplify the discussion and focus on the main physical effects, we neglect the term $(\frac{a}{a_c})^w$ hereafter, unless stated otherwise \footnote{In sections \ref{sec4} and \ref{sec5}, dust pressure is set to zero to capture the main physical features of the overdense region analytically. In the other sections, dust pressure is nonzero to avoid ambiguity in the Jeans-scale calculation.}. With this approximation, Eq.\eqref{closlefreihidjewoi} reduces to  
\bl
\left(\frac{a'}{a}\right)^2=\frac{8\pi G\rho_\text{eq}}{3}\left(\frac{a_\text{eq}^3}{a^3}+\frac{a_\text{eq}^4}{a^4}\right) a^2-\frac{1}{L^2}.\label{closlefreihi}
\el
The solution of Eq.\eqref{closlefreihi} is 
\bl
a(\eta)&=\frac{2a_\text{eq}}{\eta_\text{eq}^2/L^2}   \left[ 1-\cos \left(\frac{\eta}{L}\right)
+ \frac{\eta_\text{eq}}{L} \sin \left(\frac{\eta}{L} \right)\right],\label{hduweifiu}
\el
where
$\eta_\text{eq}\equiv\sqrt{\frac{3}{2\pi G\rho_\text{eq} a_\text{eq}^2} }$. Note that 
$a(-\eta + C L)$ is also a solution of Eq.\eqref{closlefreihi}, where $C$ is an integration constant. The solution \eqref{hduweifiu} is adopted for the overdense region because it vanishes at the initial time and reduces to the fRW scale factor in the limit 
$L\rightarrow\infty$ (see Ref. \cite{Nelson2005}). For illustration, Fig. \ref{intermorefer} shows Eq. \eqref{hduweifiu} with $\eta_\text{eq}/L=1$. The plot shows that the overdense region evolves through two phases: expansion followed by contraction, modeling the growth and collapse of the overdensity. 
The time of maximum expansion is obtained by solving  
$da(\tau)/d\tau=0$, which yields
\bl
\eta_\text{max}/L=\tan ^{-1}\left[-\frac{1}{\sqrt{1+\eta_\text{eq}^2/L^2}},\frac{\eta_\text{eq}/L}{\sqrt{1+\eta_\text{eq}^2/L^2}}\right]
=\tan^{-1}\left[-\frac{1}{\sqrt{1+\eta_{s}^2/A_\text{eq}^2}},\frac{\eta_{s}/A_\text{eq}}{\sqrt{1+\eta_{s}/A_\text{eq}}}\right], \label{definationetamax}
\el
where $\eta_s\equiv\sqrt{\frac{3}{2\pi G\rho_\text{eq}} }$ 
and $A_\text{eq}\equiv a_\text{eq} L$ for notational simplicity.  Plugging \eqref{definationetamax}
into \eqref{hduweifiu} 
and multiplying $L$
yields the maximum scale factor of the overdense region 
\bl
A_\text{max}=
\frac{2A_\text{eq}}{\eta_{s}^2/A_\text{eq}^2}\left(1+\sqrt{1+\eta_{s}^2/A_\text{eq}^2}\right). \label{dweiudhieuw}
\el
With the evolution model of the density contrast now fully specified, we proceed to discuss the threshold criteria for PBH formation in the three-zone model with dust and radiation components.

\section{Criteria for PBH Formation }\label{sec5}

In this section, we derive the conditions under which an overdense region collapses into a singularity within the cRW spacetime. Sound waves, associated with pressure, play a crucial role in resisting gravitational collapse. Taking sound propagation into account, there must exist a minimum value of the boundary coordinate $\chi_a$, the smallest radius of the overdense region, for PBH formation to occur. Otherwise, sound waves have sufficient time to propagate throughout the region and significantly suppress collapse. We employ two criteria to quantify this effect: (i) comparing the sound propagation timescale with the timescale of maximum expansion (Criterion 1);
(ii) comparing the sound propagation timescale with the timescale of apparent horizon formation (Criterion 2).    
The minimum boundary coordinate $\chi_a$ can then be determined under these two criteria. The formalism relevant to these two criteria was originally developed for a single‑fluid system in Ref.~\cite{Harda2013}, here we generalize it to a two-fluid system consisting of dust and radiation.

\subsection{Criterion 1: Comparison of sound‑wave propagation and maximum‑expansion times} \label{C1}

Ref.~\cite{Harda2013} argued that black holes can form if the time required for a sound wave to propagate from the centre to the surface of the overdense region exceeds the time of maximum expansion given by Eq.~\eqref{definationetamax}. In this case, the sound‑propagation timescale is longer than the gravitational evolution timescale, so pressure does not have sufficient time to significantly modify the evolution of the overdense region.

The background quantum bouncing model contains both expanding and contracting phases. During the background expanding phase, the full evolution of the cRW scale factor in Eq.~\eqref{hduweifiu} must be taken into account, since the overdense region is expected to co‑expand with the background universe before eventually collapsing~\cite{Carr2025}. In the background contracting phase, the overdense region contracts together with the background universe and, if accretion is neglected, does not undergo a prior expansion. For Criterion 1, the durations of the expanding and contracting phases are equal in the cRW spacetime, so the conditions for PBH formation are identical in both phases.

Criterion 1 can be written as 
\bl
\eta_\text{max}\leq \eta_\text{sound}(\chi_a),\label{confition}
\el
where $\eta_\text{sound}(\chi_a)$ denotes the 
travel time of a sound wave from the centre ($\chi=0$) of the overdense region to its surface ($\chi=\chi_a$).
The sound-wave equation is \cite{Harda2013}
\bl
\frac{ d\chi}{ d (\eta/L)}&=c_s,\label{soundrqu}
\el
where $c_s$ is the sound speed, defined by 
\bl
c_s^2
\equiv\frac{d(p_{w}+p_{r})/d\eta}{d(\rho_{w}+\rho_{r})/d\eta}
=\frac13\frac{d\rho_{r}/d\eta}{d(\rho_{w}+\rho_{r})/d\eta},\label{eqcs}
\el
with $\rho_w$ and $\rho_r$ the energy densities of 
dust and radiation, respectively, and $p_w=0,~p_r=1/3\rho_r$. Using the conservation equations  
\begin{subequations}
\bl
\frac{d\rho_w}{d \eta}+3{\cal H}\rho_w&=0,
\\
\frac{d\rho_r}{d\eta}+4{\cal H}\rho_r&=0,
\el
\end{subequations}
together with $\rho_w/\rho_\text{eq}=(a_\text{eq}/a)^3$
and $\rho_r/\rho_\text{eq}=(a_\text{eq}/a)^4$, one has   
\bl
c_s^2
=\left((9/4)(a/a_\text{eq})+3\right)^{-1}.\label{css}
\el
Eq.~\eqref{css} also holds in the background FRW spacetime when $(\bar{a}/\bar{a}_c)^w$ approaches unity, since \(w\) is extremely small. Plugging 
Eqs. \eqref{hduweifiu} and \eqref{css} 
into the right-hand side of Eq. \eqref{soundrqu} and integrating yields 
the equation of the  wave front 
{\footnotesize \bl
\chi&=\frac{2\eta_{s}/A_\text{eq}}{\sqrt{\frac{9}{2}+3\frac{\eta_{s}^2}{A_\text{eq}^2}
+\frac{9}{2}\sqrt{1+\frac{\eta_{s}^2}{A_\text{eq}^2}}}}\left\{F\left[\frac12(\frac{\eta}{\eta_\text{eq}}\frac{\eta_s}{A_\text{eq}}-\frac{\eta_\text{max}}{L}) \Big|\frac{\sqrt{1+\eta_{s}^2/A_\text{eq}^2}}{\frac{1}{2}+\frac{\eta_{s}^2/A_\text{eq}^2}{3}
+\frac{1}{2}\sqrt{1+\eta_{s}^2/A_\text{eq}^2}}\right] -F\left[-\frac12\frac{\eta_\text{max}}{L}\Big|\frac{\sqrt{1+\eta_{s}^2/A_\text{eq}^2}}{\frac{1}{2}+\frac{\eta_{s}^2/A_\text{eq}^2}{3}
+\frac{1}{2}\sqrt{1+\eta_{s}^2/A_\text{eq}^2}}\right] \right\}\label{newfhuirewfheuirnew}
\el}
where $F(a|b)$ is the elliptic integral of the first kind. The detailed calculation is presented in Appendix~\ref{Appa}.

\begin{figure}[htb]
 \subcaptionbox{}
   {%
     \includegraphics[width = .45\linewidth]{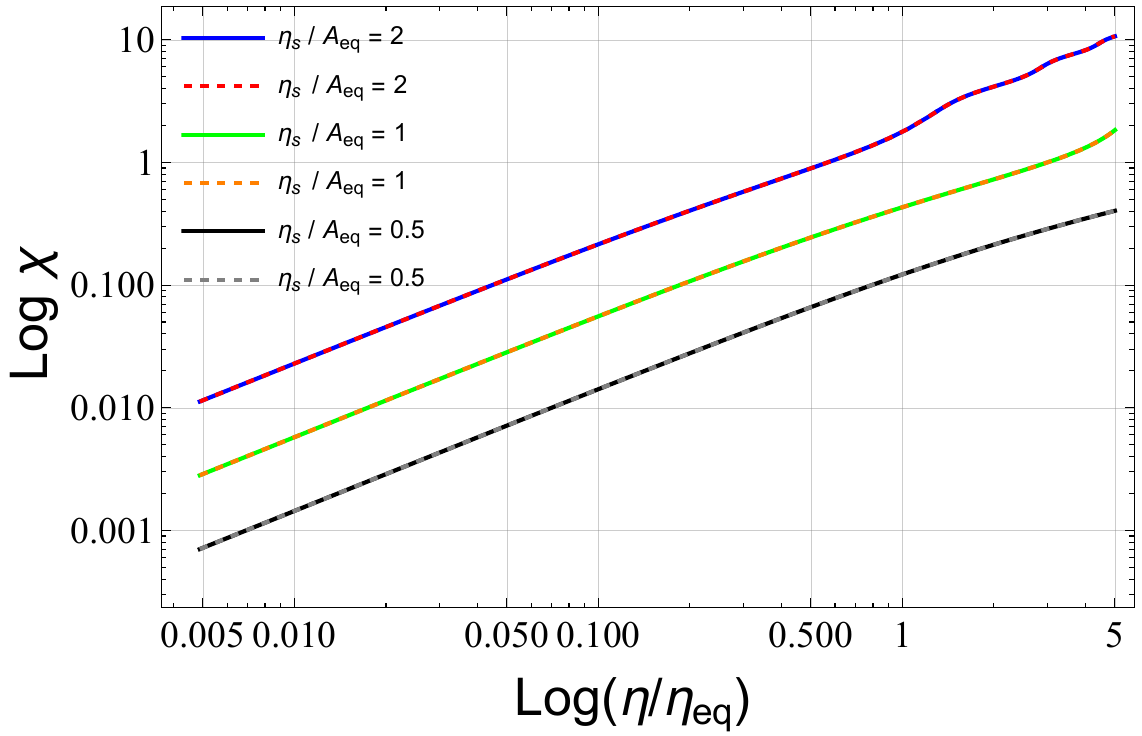}}
 \centering
 \hspace{0.5cm}
 \subcaptionbox{}
   {%
     \includegraphics[width = .35\linewidth]{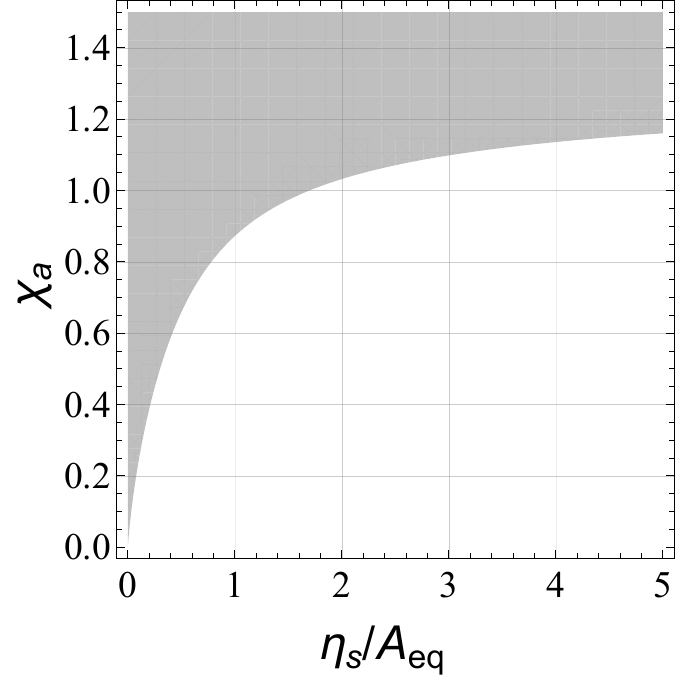}}
\caption{
(a) Wave‑front position of the sound wave as a function of time. Solid (thick) line: analytic solution of Eq. \eqref{soundrqu}, dashed line: numerical integration of Eq.\eqref{soundrqu}.
(b) Shaded region: Values of $\chi_a$ and $\eta_s/ A_\text{eq}$ for which PBH formation occurs.} \label{chi12}
\end{figure}
As a consistency check, we plot both the analytic solution (solid line) and the numerical solution (dashed line) of Eq. \eqref{soundrqu} for parameters $\eta_\text{eq}/L=\eta_{s}/A_\text{eq}= 0.5, 1, 2$
in Fig.\ref{chi12} (a). The two solutions agree well. The results show that 
the sound speed increases with $\eta_{s}/A_\text{eq}$, since a sound wave with larger $\eta_{s}/A_\text{eq}$ can propagate a greater comoving distance at a given time.
The inverse function of Eq.\eqref{newfhuirewfheuirnew} is 
{\footnotesize
\bl
\frac{\eta}{L}=\frac{\eta_\text{max}}{L}+2 \text{JA}\left\{\frac{\sqrt{6}}{4}\frac{A_\text{eq}}{\eta_{s}}
\sqrt{2+\frac{\eta_{s}^2}{A_\text{eq}^2}+3\sqrt{1+\eta_s^2/A_\text{eq}^2}} \chi-F\left[\frac12\frac{\eta_\text{max}}{L}|\frac{\sqrt{1+\eta_{s}^2/A_\text{eq}^2}}{\frac{1}{2}+\frac{\eta_{s}^2/A_\text{eq}^2}{3}
+\frac{1}{2}\sqrt{1+\eta_{s}^2/A_\text{eq}^2}}\right],\frac{\sqrt{1+\eta_{s}^2/A_\text{eq}^2}}{\frac{1}{2}
+\frac{\eta_{s}^2/A_\text{eq}^2}{3}
+\frac{1}{2}\sqrt{1+\eta_{s}^2/A_\text{eq}^2}}\right\},
\label{dguwyergfiunew}
\el}
where JA denotes the JacobiAmplitude function \cite{NIST:DLMF}.
Plugging Eq. \eqref{dguwyergfiunew} into the criterion \eqref{confition} yields a lower bound on the boundary coordinate 
$\chi_a$
\bl
\chi_a\geq\frac{\sqrt{8/3}\eta_{s}/A_\text{eq}}{\sqrt{3+\eta_s^2/A_\text{eq}^2+3\sqrt{1+\eta_s^2/A_\text{eq}^2}}}
F\left[\frac12\frac{\eta_\text{max}}{L}|\frac{\sqrt{1+\eta_s^2/A_\text{eq}^2}}{\frac{1}{2}+\frac{\eta_s^2/A_\text{eq}^2}{3}
+\frac{1}{2}\sqrt{1+\eta_s^2/A_\text{eq}^2}}\right],\label{218hfueritext}
\el
which depends only on the ratio $\eta_s/A_\text{eq}$. Therefore, for a given  $\eta_s/A_\text{eq}$, one can determine a threshold value of the surface coordinate. If 
$\chi_a$ lies below this threshold, pressure has sufficient time to prevent PBH formation. The relation in Eq.~\eqref{218hfueritext} is plotted in Fig.~\ref{chi12}(b), where PBH formation is allowed in the shaded region.

\subsection{Criterion 2: Compare the sound‑wave propagation time with the apparent‑horizon formation time} \label{subsec5p2}

Ref.~\cite{Harda2013} proposed a stricter criterion for PBH formation than Criterion 1. Specifically, the time required for a sound wave to propagate between the centre and the surface of the overdense region (or vice versa) 
\begin{figure}[htb]
 \subcaptionbox{}
   {%
     \includegraphics[width = .45\linewidth]{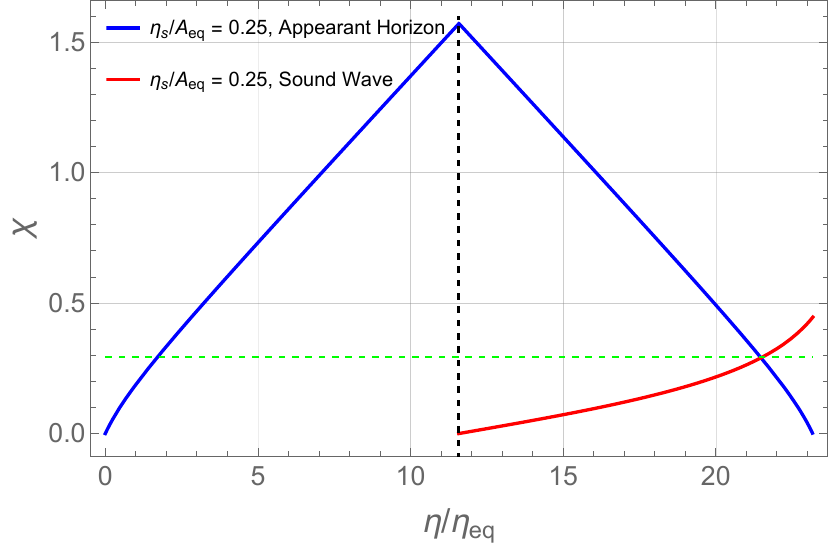}}
 \centering
 \subcaptionbox{}
   {%
     \includegraphics[width = .45\linewidth]{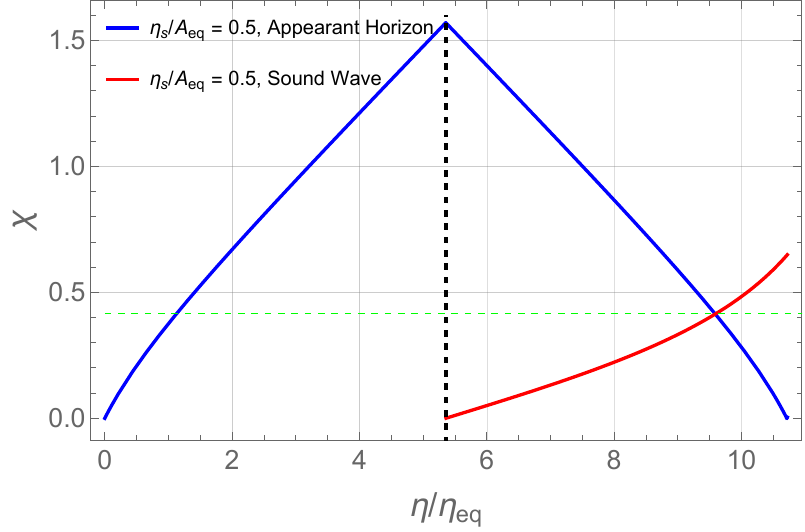}}
 \subcaptionbox{}
   {%
     \includegraphics[width = .45
\linewidth]{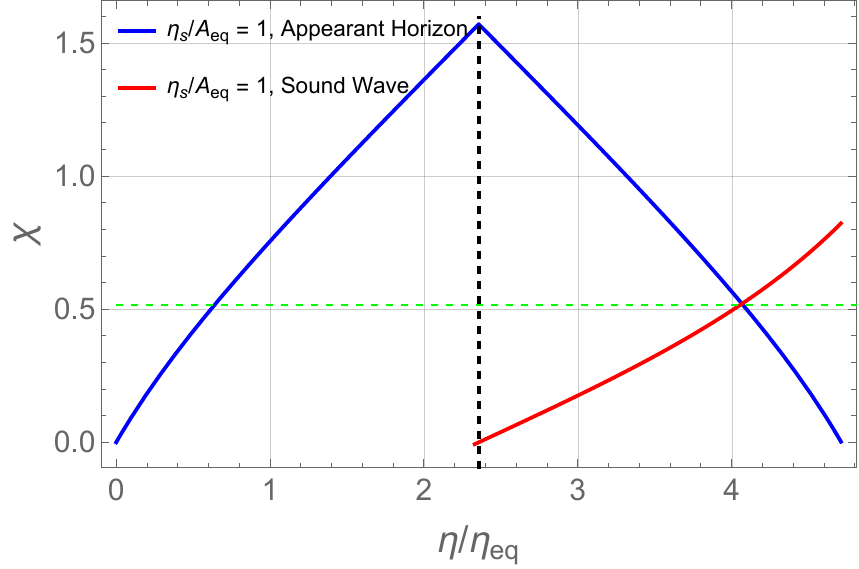}} 
\subcaptionbox{}
   {%
     \includegraphics[width = .45
\linewidth]{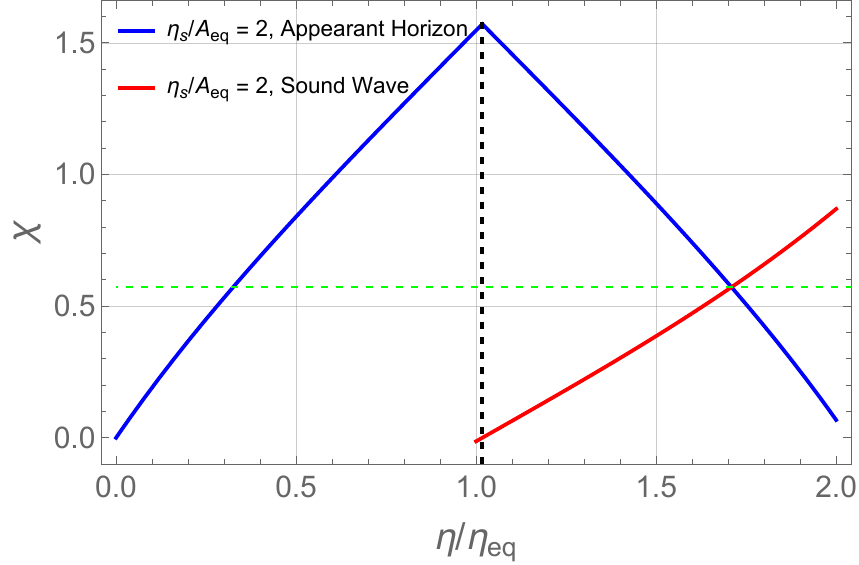}} 
\caption{Contracting background universe. Blue curve: Left and right halves represent the past and future apparent horizons, 
respectively.
Red curve: Sound wave front.} \label{ccfrefgerThorZion}
\end{figure}
must exceed the time at which an apparent horizon forms. When the sound‑propagation timescale is longer than the apparent‑horizon formation timescale, pressure forces cannot act effectively to prevent collapse, and a black hole is expected to form.

Unlike Criterion 1, Criterion 2 yields two distinct thresholds for PBH formation depending on whether the background is contracting or expanding. In the contracting background, the initial time of the cRW patch is chosen at the moment of maximum expansion, so only the subsequent collapse phase of the overdense region is relevant. By contrast, in the expanding background, the full cRW evolution must be considered, since the overdense region is expected to co‑expand with the background universe prior to collapse~\cite{Harda2013}.

Since we focus primarily on the 
PBH formation in a contracting background, in this subsection we present an explicit calculation of the 
lower bound on $\chi_a$ for the contracting case. 
The corresponding calculation for the expanding background will be sketched briefly for completeness.  

To evaluate the apparent-horizon formation time, we use the 
the Minser-Sharp mass ${\cal M}$ for a cRW patch \cite{Hayward, Misner, Harda2013}
\bl
{\cal M}=\frac{a L}{2G}\left[1+\frac{(da/d(\eta/L))^2}{a^2}\right]\sin^3\chi.
\el
The apparent horizon is located where $\frac{2GM}{R}=
\frac{2GM}{aL\sin\chi}=1$ \cite{Harda2013}, which yields the condition  
\bl
\left(1+L^2\frac{a'^2}{a^2}\right)\sin^2\chi=1.\label{hdiuh}
\el
Plugging the cRW scale factor \eqref{hduweifiu} in Eq. \eqref{hdiuh} and solving for $\chi$ gives  
\bl
\chi=\arcsin\left[\left(1+\frac{\big(\frac{\eta_s}{A_\text{eq}}\cos(\frac{\eta}{\eta_\text{eq}}\frac{\eta_s}{A_\text{eq}})+\sin(\frac{\eta}{\eta_\text{eq}}\frac{\eta_s}{A_\text{eq}})\big)^2}{\big(1-\cos(\frac{\eta}{\eta_\text{eq}}\frac{\eta_s}{A_\text{eq}})+\frac{\eta_s}{A_\text{eq}}\sin(\frac{\eta}{\eta_\text{eq}}\frac{\eta_s}{A_\text{eq}})\big)^2}\right)^{-1/2}\right],\label{dhewuihfdui}
\el
which gives the apparent-horizon radius as a function of time.  Although four branches of solutions exist, two of them correspond
$\chi\in (\pi, 2\pi)$ and lie outside the range of Type I regions considered in this work (see Eq. \eqref{eq37}). Among the remaining two, one solution is not suitable for discussing PBH formation, as the sound-wave front does not intersect the apparent horizon. This implies that the all overdense region can collapse into black hole, which is unphysical. Therefore, only the solution given in Eq.~\eqref{dhewuihfdui} is physically relevant for the formation of PBH.

We plot the evolution of the apparent horizon (blue curve), given by Eq.~\eqref{dhewuihfdui}, together with the wavefront of the sound wave (red curve),
\bl
\chi&=\frac{2\eta_{s}/A_\text{eq}}{\sqrt{\frac{9}{2}+3\frac{\eta_{s}^2}{A_\text{eq}^2}
+\frac{9}{2}\sqrt{1+\frac{\eta_{s}^2}{A_\text{eq}^2}}}}F\left[\frac12\left(\frac{\eta}{\eta_\text{eq}}\frac{\eta_s}{A_\text{eq}}-\frac{\eta_\text{max}}{L}\right) |\frac{\sqrt{1+\eta_{s}^2/A_\text{eq}^2}}{\frac{1}{2}+\frac{\eta_{s}^2/A_\text{eq}^2}{3}
+\frac{1}{2}\sqrt{1+\eta_{s}^2/A_\text{eq}^2}}\right],\label{newfhuirewfheuirnew1}
\el
and show both in Fig.~\ref{ccfrefgerThorZion}(a)–(d) for parameter choices $\eta_s/A_\text{eq}=0.25, 0.5, 1, 2$. Note that Eq. \eqref{newfhuirewfheuirnew1} differs from \eqref{newfhuirewfheuirnew} because, in the contracting background, the evolution of the density contrast is taken to begin at the time of maximal expansion.

The left and right halves of the blue curves in Fig.~\ref{ccfrefgerThorZion} denote the past and future apparent horizons, respectively. The green dashed horizontal lines mark the intersections between the sound‑wave fronts (red) and the future apparent horizon, which determine the lower bound on the surface coordinate $\chi_{a}$.  For the contracting background (panels (a)–(d), with $\eta_s/A_\text{eq}=0.25, 0.5, 1, 2$), the corresponding lower bounds
$\chi_a$ are listed in the second row of Table. \ref{tab}. For completeness, we also compute the lower bounds $\chi_a$ in the expanding background. In this case, the sound-wave front is given by Eq.~\eqref{newfhuirewfheuirnew} (rather than Eq.~\eqref{newfhuirewfheuirnew1}), and the same procedure yields the results shown in the third row of Table~\ref{tab}.
\begin{table}[htbp] 
\centering \begin{tabular}{|c|c|c|c|c|} \hline \diagbox{$\chi_a$}{$\eta_s/A_{\rm eq}$} & 0.25 & 0.5 & 1.0 & 2.0 \\ \hline Contracting background & 0.292 & 0.416 & 0.515 & 0.574 \\ \hline Expanding background & 0.624 & 0.862 & 1.062 & 1.145 \\ \hline \end{tabular} \caption{Values of $\chi_a$ for different $\eta_s/A_{\rm eq}$ within contracting and expanding backgrounds.} \label{tab} 
\end{table}

\begin{figure}[htb]
 \centering
   {%
     \includegraphics[width = .5\linewidth]{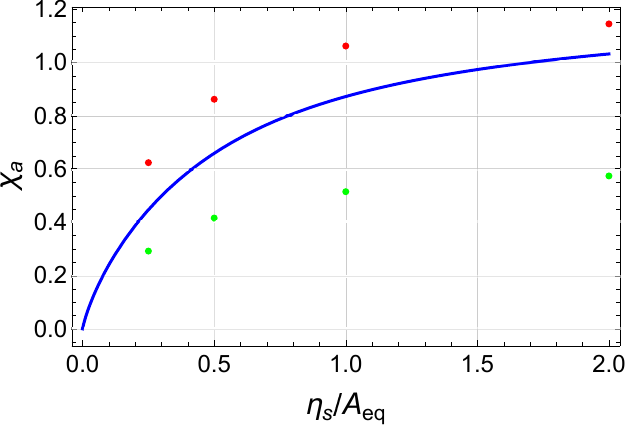}}
\caption{Lower bounds on the surface coordinate $\chi_a$
 from Criterion 1 and Criterion 2. Blue curve: Criterion 1, applicable to both expanding and contracting backgrounds.
Red dots: Criterion 2 in the expanding background.
Green dots: Criterion 2 in the contracting background.} \label{dfrefgerThorZion}
\end{figure}

To compare the lower bounds on the surface coordinate $\chi_a$ implied by Criterion 1 and Criterion 2, we plot 
Fig. \ref{dfrefgerThorZion} for illustration. The figure shows that, in an expanding background, the lower bound from Criterion 2 (red dots) is generally larger than that from Criterion 1 (blue line), whereas in a contracting background the lower bound from criterion 2  (green dots) is 
generally smaller than that from Criterion 1. 
Thus, Criterion 2 allows a broader range of perturbation sizes capable of forming  PBHs in the contracting background. In other words, PBH formation is easier during the contraction than during the expanding background, consistent  with physical intuition. Fig. \ref{dfrefgerThorZion} also shows that the discrepancy between the two criteria grows as $\eta_{s}/A_\text{eq}$ increases and vanishes in the limit $\eta_{s}/A_\text{eq}\rightarrow0$. This point will be further discussed in Sect. \ref{sec6}, where we focus exclusively on the PBH formation in the contracting background.

\section{PBH Formation in Contracting Background}\label{sec6}

In this section we derive the mass fraction $\beta$ of PBHs of a given mass
at formation time $(\bar{a}_{\text{Form}}/\bar{a}_c)$. The quantity $\beta$ represents the fraction of matter perturbations that collapse into PBHs, and it is an observable constrained by current data \cite{1974ApJ18425P, 1975ApJ2011C}. Following Ref. \cite{Harda2013}, the mass fraction $\beta$ is defined as
\bl
{\beta =\int_{\zeta_c}^{\infty}\frac{2}{\sqrt{2\pi {\cal P}_{\zeta}(k_{J},\bar{a}_{\text{Form}}/\bar{a}_c)}}
\text{exp}\left\{-\frac{\zeta^2}{2{\cal P}_{\zeta}(k_{J},\bar{a}_{\text{Form}}/\bar{a}_c)}\right\}d\zeta }=
\text{Erfc}\left[\frac{\zeta_c}{\sqrt{2{\cal P}_{\zeta}(k_J,\bar{a}_{\text{Form}}/\bar{a}_c)}}\right],\label{massfraction}
\el
where $\zeta$ is the curvature perturbation, assumed to follow a Gaussian distribution, and  $P_{\zeta}(k_J,\bar{a}_{\text{Form}}/\bar{a}_c)$
is the curvature power spectrum  evaluated at the Jeans wavenumber $k_J$ and at the PBH formation time $(\bar{a}_{\text{Form}}/\bar{a}_c)$. 
The threshold $\zeta_c$ is the critical curvature perturbation above which collapse to a PBH occurs. Erfc is the complementary error function.

We assume that PBH formation occurs when the perturbation mode with comoving wavenumber $k$ crosses the Jeans scale $k_J$. The PBH mass is taken to be the mass contained within the Jeans length $R_J\equiv\frac{2\pi}{k_J}$ (where $k_J$ is defined in Eq. \eqref{kJcomi}), including contributions from both background and perturbed energy densities. At the formation time, one has 
\bl
M_J=\frac{4\pi}{3} (\bar{\rho}+\delta \rho_{k=k_J})\Big(
\frac{R_J\bar{a} }{2}\Big)^3\approx \frac{4\pi}{3}(\bar{\rho}_{r}+\bar{\rho}_{w})\Big(\frac{\bar{a}R_J}{2}\Big)^3,\label{dwedioew}  
\el
where  
$\bar{\rho}=\bar{\rho}_{w}+\bar{\rho}_r$ and  $\delta \rho_k=\delta\rho_{w,k}+\delta\rho_{r,k}$.
The factor $\bar{a}/2$ converts the comoving Jeans length into the physical radius, and we take the PBH radius to be half the physical Jeans length. In the perturbative regime
$\delta\rho_k\ll\bar{\rho}$, so for estimating the formation time it is convenient to neglect
$\delta \rho_k$, as shown in the second equality of Eq.\eqref{dwedioew}.  Dropping 
$\delta \rho_k$ simplifies the estimate but does not imply that the PBH mass in reality lacks contribution from local perturbations. 

Plugging $R_J\equiv 2\pi/k_J$ together with Eq.\eqref{kJcomi} into Eq. \eqref{dwedioew} yields 
\bl
\frac{M_{J}}{M_{\ensuremath{\odot}}}=
 10^{23}\times
\frac{8 \sqrt{2} \pi ^3 w^{3/2}\Big(\Omega_{w, c}(\bar{a}_c/\bar{a})^3+ \Omega_{r, c}(\bar{a}_c/\bar{a})^4\Big)}{\left(6\Omega_{w, c}(\frac{\bar{a}_c}{\bar{a}})^3-(1-48 w) \Omega_{r, c}(\frac{\bar{a}_c}{\bar{a}})^4+ \sqrt{[6\Omega_{w, c}(\frac{\bar{a}_c}{\bar{a}})^3-(1-48 w) \Omega_{r, c}(\frac{\bar{a}_c}{\bar{a}})^4]^2+192 w \Omega_{r, c}^2(\frac{\bar{a}_c}{\bar{a}})^8}\right)^{3/2}},\label{dhewui}
\el
where we have used $1/(G H_0)\approx  10^{23} M_{\ensuremath{\odot}}$ and Eq. \eqref{sothat}. In addition, we 
set $(\bar{a}_c/\bar{a})^{3w}\rightarrow 1$, since $w\approx 9.440 \times10^{-22}\approx0$, which simplifies the calculation in Eq. \eqref{dhewui}.

Plugging the four masses in  Table. \ref{tablm1} into the left-hand side of Eq. \eqref{dhewui} yields the corresponding formation times. The Jeans scales at formation are then obtained by substituting these formation times into Eq. \eqref{kJcomi}. The results are summarized in Table \ref{table1N},
\begin{table}[h]
\begin{tabular}{|c|c|c|c|c|}
\hline
& $\bar a_{\text{Form}}/\bar{a}_c$ & $x_{\text{Form}}$ & $k_J$(s$^{-1}$) & ${\cal P}_{\zeta}(k_J,\frac{\bar a_{\text{Form}}}{\bar{a}_c})$ \\
\hline
$10^{13}\,M_\odot$  & $5.506\times10^{13}$ & $-1.484\times 10^{7}$ & $7.554 \times 10^{-15}$ & $2.826 \times 10^{-152}$\\
\hline
$10^{2}\,M_\odot$ & 
$2.556 \times 10^{6}$ & $-3.197\times 10^3$ & 
$3.506 \times 10^{-11}$ & $5.983\times 10^{-127}$\\
\hline
$10^{-14}\,M_\odot$ & $5.494\times10^{-5}$ & $-1.408\times 10^{-2}$ &
$7.561 \times 10^{-6}$ & $1.296\times 10^{-89}$ \\
\hline
$10^{-17}\,M_\odot$ & $4.771\times10^{-7}$ & $-8.159 \times10^{-4}$ & 
$7.907 \times 10^{-5}$  & $1.581\times 10^{-82}$ \\
\hline
\end{tabular}
\centering
\caption{Mapping between PBH mass and related quantities: formation time, comoving Jeans wavenumber and the curvature perturbation spectrum.}\label{table1N}
\end{table}
while the curvature power spectra corresponding to the masses listed 
in Table \ref{table1N} are shown in Figs. \ref{powerspectrumofiin} (a) - (d). In Sect. \ref{sec3} we showed that the Jeans scale is no longer valid after the transition scale factor $\bar{a}_\text{Tr}/\bar{a}_c\approx 10^{-8}\approx 0.1\times(\bar{a}_\text{eq}/\bar{a}_c)$. 
From the second column of Table \ref{table1N}, all scale factors at PBH formation lie before the transition scale factor. 
For a given PBH mass, Eq. \eqref{dhewui} generally admits multiple solutions for the formation scale factor  
$\bar{a}_{\text{Form}}/\bar{a}_c$. However, if we require that $\bar{a}_{\text{Form}}$ is real and satisfies $\bar{a}_{\text{Form}}>\bar{a}_\text{Tr}$, then the solution for $\bar{a}_{\text{Form}}/\bar{a}_c$ is unique.

The mass fraction $\beta$ in Eq. \eqref{massfraction} depends on the threshold curvature perturbation, $\zeta_c$, which  is determined by the choice of collapse model. In the three-zone model, the critical curvature perturbation appearing in Eq. \eqref{massfraction}, $\zeta_c$, is given by \cite{PhysRevD.83.124025}
\bl
\zeta_c\approx -2\ln \cos\frac{\chi_{a}(\eta_s,A_\text{eq})}{2},\label{fheurifh}
\el
which depends on the surface coordinate $\chi_a$, itself a  function of the parameters $\eta_s$ and $A_\text{eq}$ as described in Sect. \ref{sec5}. For the parameter ratios $\eta_s/A_\text{eq}=0.25, 0.5, 1, 2$, we derived lower bounds on $\chi_a$ using Criterion 1 (see \eqref{218hfueritext}) and Criterion 2 (see Table. \ref{tab}). Substituting those bounds into Eq. \eqref{fheurifh} gives corresponding estimates for $\zeta_c$. However, when PBH formation is considered in the quantum-bouncing universe, the parameters $\eta_s$ and $A_\text{eq}$ for an overdense region are not free, their values are fixed by  
matching the three-zone model to the quantum bouncing model.

To determine the two quantities $\eta_s$ and $A_\text{eq}$
for a PBH of given mass, we adopt two assumptions: (i)  collapse occurs immediately when the overdense scale  
crosses the Jeans scale, and (ii) collapse initiates for the cRW patch at its time of maximum 
expansion \cite{PhysRevD.83.124025}. Under these assumptions, we obtain the following two coupled  algebraic equations, 
\begin{subequations}\label{coupledrquation}
\bl
\frac12 R_{J}(\bar{a}_{\text{Form}})\bar{a}_{\text{Form}}&=A_\text{max}  \sin\chi_{a},\label{frejiof}
\\
\bar{\rho}(\bar{a}_{\text{Form}})+\delta\rho_{k=k_J}(\bar{a}_{\text{Form}}&)\approx \bar{\rho}(\bar{a}_{\text{Form}}) =\rho_\text{max},\label{dewhduiwe}
\el
\end{subequations}
where the left-hand sides of Eqs. \eqref{coupledrquation}
correspond to quantities in the quantum bouncing model. In Eq. \eqref{dewhduiwe}, we have used the approximation 
$\delta\rho_{k=k_J}\ll\bar{\rho}$ in the second equality, consistent with perturbation theory.
The right-hand sides of Eqs. \eqref{coupledrquation}
involve the cRW quantities discussed in Sect.
\ref{sec4}. The energy density of the cRW at maximum expansion, $\rho_\text{max}$, is given by 
\bl
\frac{\rho_\text{max}}{3/(2\pi G)}=\frac{\eta_{s}^4}{8A_\text{eq}^6}\left(1+\sqrt{1+\frac{\eta_{s}^2}{A_\text{eq}^2}}\right)^{-3}\left[1+\frac{\eta_{s}^2}{2A_\text{eq}^2}\left(1+\sqrt{1+\frac{\eta_{s}^2}{A_\text{eq}^2}}\right)^{-1}\right],\label{dwedhiu}
\el
where we used $\rho_\text{max}=\rho_\text{eq}(a_\text{eq}/a_\text{max})^{3}(1+a_\text{eq}/a_\text{max})$, together with 
Eq. \eqref{dweiudhieuw} and $\rho_\text{eq}=3/(2\pi G\eta_s^2)$. The two equations in \eqref{coupledrquation} are 
sufficient to solve for the two unknowns $\eta_s$ and $A_\text{eq}$. 

With these equations, we can calculate $\eta_s$ and $A_\text{eq}$ by solving Eqs. \eqref{coupledrquation}.  From the solutions we then obtain the critical curvature perturbation and the mass fraction $\beta$. In Sects. \ref{sub61} and \ref{sub62} we perform these calculations for each of the criteria introduced in Sects.~\ref{C1} and \ref{subsec5p2}.

\subsection{Mass fraction from Criterion 1} 
\label{sub61}

As an example, we illustrate the calculation of the critical 
curvature perturbation for a PBH with mass 
$10^{13}\,M_\odot$. The same procedure applies to other masses, such as $10^2, 10^{-14}, 10^{-17} \,M_\odot$.

Plugging the expression for the maximum scale factor from \eqref{dweiudhieuw} and the lower bound of the surface coordinate from criterion 1 \eqref{218hfueritext} into  
Eq. \eqref{frejiof} yields 
{\footnotesize 
\bl
\frac14 R_{J}(\bar{a}_{\text{Form}})\bar{a}_{\text{Form}}
&=\frac{A_\text{eq}^3 }{\eta_s^2}(1+\sqrt{1+\eta_{s}^2/A_\text{eq}^2})
\sin\left[\frac{\sqrt{8/3}}{\sqrt{3\frac{A_\text{eq}^2}{\eta_s^2}+1+3\frac{A_\text{eq}^2}{\eta_s^2}\sqrt{1+\frac{\eta_s^2}{A_\text{eq}^2}}}}
F\left[\frac12\frac{\eta_\text{max}}{L}\Bigg|\frac{\sqrt{A_\text{eq}^2/\eta_s^2+1}}{\frac{1}{2}\frac{A_\text{eq}}{\eta_s}
+\frac{\eta_s}{3A_\text{eq}}
+\frac{1}{2}\frac{A_\text{eq}}{\eta_s}\sqrt{1+\frac{\eta_s^2}{A_\text{eq}^2}}}\right]\right],\label{dhweuidhe}
\el}
where the Jeans length  $R_J\equiv 2\pi/k_J$ can be obtained from Eq. \eqref{kJcomi}. Plugging 
Eq. \eqref{dwedhiu}, together with $\bar{\rho}(\bar{a}_{\text{Form}})=\frac{3\bar{H}^2_0}{8\pi G }[\Omega_{w, c}(\bar{a}_c/\bar{a}_{\text{Form}})^3+ \Omega_{r,c}(\bar{a}_c/\bar{a}_{\text{Form}})^{4}]$
into Eq. \eqref{dewhduiwe} yields 
\bl
\frac{H^2_0}{4}\left[\Omega_{w,c}\left(\frac{\bar{a}_{c}}{\bar{a}_{\text{Form}}}\right)^3+ \Omega_{r, c}\left(\frac{\bar{a}_{c}}{\bar{a}_{\text{Form}}}\right)^4\right]
=\frac{\eta_{s}^4}{8A_\text{eq}^6}\left(1+\sqrt{1+\frac{\eta_{s}^2}{A_\text{eq}^2}}\right)^{-3}\left[1+\frac{\eta_{s}^2}{2A_\text{eq}^2}\left(1+\sqrt{1+\frac{\eta_{s}^2}{A_\text{eq}^2}}\right)^{-1}\right]. \label{dewhduiwe22}
\el
Using $(\bar{a}_{\text{Form}} / \bar{a}_{c})=5.506\times10^{13} $ (see Table \ref{table1N}) and solving Eqs. 
\eqref{dhweuidhe} and \eqref{dewhduiwe22} simultaneously, we find $(A_\text{eq},  \eta_s)=(1.541 \times 10^{17} \text{s}, 8.978 \times 10^6 \text{s})$. Substituting these values into the lower bound for the surface coordinate from criterion 1, Eq. \eqref{218hfueritext}, yields $\chi_a= 7.881\times 10^{-11}$. Inserting 
$\chi_a$ into Eq. \eqref{fheurifh} then yields the 
critical curvature perturbation $\zeta_c= 1.553 \times 10^{-21}$ for a
PBH for mass $10^{13} M_\odot$. The set $(A_\text{eq}, \eta_s, \chi_a, \zeta_c)$ for different PBH masses is summarized in Table \ref{table1}.
\begin{table}[h]
\begin{tabular}{|c|c|c|c|c|}
\hline
\textbf{Criterion 1} &  $A_{\rm eq}\,(\mathrm{s})$ & $\eta_s\,(\mathrm{s})$ & $\chi_a$ & $\zeta_c$  \\
\hline
$10^{13}\,M_\odot$ & $1.541\times 10^{17}$ & $8.978\times10^{6}$  & $7.881\times10^{-11}$ & $1.553\times10^{-21}$  \\
\hline
$10^{2}\,M_\odot$ & $1.541\times10^{6}$ & $8.977\times10^{-5}$ & $7.880\times10^{-11}$ & $1.552\times10^{-21}$ \\
\hline
$10^{-14}\,M_\odot$ & $1.539\times10^{-10}$ & $8.982\times10^{-21}$ & $7.894\times10^{-11}$ & $1.558\times10^{-21}$ \\
\hline
$10^{-17}\,M_\odot$ & $1.495\times10^{-13}$ & $1.021\times 10^{-23}$ & 
$9.236\times 10^{-11}$&
$2.133\times 10^{-21}$\\
\hline
\end{tabular}
\centering
\caption{Values of $(A_\text{eq}, \eta_s, \chi_a, \zeta_c)$ obtained from criterion 1  for PBHs of different masses. }
\label{table1}
\end{table}

An interesting observation is that 
the critical curvature perturbation $\zeta_c$ appears to be nearly mass-independent over a large mass range, $10^{-14}
\sim 10^{13} M_\odot$ (see the $\zeta_c$ column in Table \ref{table1}). This behavior arises because the ratio $\eta_s/A_\text{eq}$ takes approximately the same value for masses $10^{13}, 10^{2}, 10^{-14} M_{\ensuremath{\odot}}$
\bl
\frac{\eta_s}{A_\text{eq}} \approx 5.826\times 10^{-11},\label{dewjio}
\el
and both $\chi_a$ and $\zeta_c$ depend only on this ratio. Note that the surface coordinate 
$\chi_a$ being identical for these masses does not imply that the physical radius of PBHs are the same, as the physical radius is  
\bl
A_\text{max}  \sin\chi_a=2A_\text{eq} \frac{ A_\text{eq}^2}{\eta_{s}^2}\left(1+\sqrt{1+\frac{\eta_{s}^2}{A_\text{eq}^2}}\right) \sin\chi_a\left(\frac{\eta_{s}}{A_\text{eq}}\right),\label{eq71w}
\el
where Eq.\eqref{dweiudhieuw} has been used. 
Eq. \eqref{eq71w} shows that the PBH radius is proportional to  $A_\text{eq}$ in the 
mass range  $10^{-14}
\sim 10^{13} M_\odot$,
which decreases with the decreasing PBH mass, see $A_\text{eq}-$column in Table \ref{table1}.

In what follows, we show analytically that $\eta_s/A_\text{eq}$ is identical for PBH masses $10^{13}, 10^{2}, 10^{-14} M_{\ensuremath{\odot}}$ and depends only on the 
equation of state $w$. Consequently, both $\chi_a$ and 
$\zeta_c$ depend only on $w$. To simplify notation, define 
\begin{subequations}
    \bl
F\left(\frac{\eta_s}{A_\text{eq}}\right)&\equiv \frac{\eta_s^2 }{A_\text{eq}^3}\frac14 R_{J}(\bar{a}_{\text{Form}})\bar{a}_{\text{Form}},\label{eq71}
\\
G\left(\frac{\eta_s}{A_\text{eq}}\right)&\equiv \frac{ A_\text{eq}^6}{\eta_s^4}\frac{\bar{H}^2_0}{4}\Big(\Omega_{w,c}\left(\frac{\bar{a}_{c}}{\bar{a}_{\text{Form}}}\right)^3+ \Omega_{r, c}\left(\frac{\bar{a}_{c}}{\bar{a}_{\text{Form}}}\right)^4\Big),\label{eq72}
\el
\end{subequations}
where Eqs. \eqref{dhweuidhe} and \eqref{dewhduiwe22} have been used. 

Squaring Eq. \eqref{eq71}, multiplying by Eq. \eqref{eq72}, substituting $R_J=2\pi/k_J$, using Eq.\eqref{kJcomi}, and performing a small $w$ expansion 
yields 
\bl
F\left(\frac{\eta_s}{A_\text{eq}}\right)^2G\left(\frac{\eta_s}{A_\text{eq}}\right)\approx w\frac{\pi^2}{4}\frac{1+\bar{a}_c/\bar{a}_{\text{Form}}\times\Omega_{r,c}/\Omega_{w,c}}{6-\bar{a}_c/\bar{a}_{\text{Form}}\times\Omega_{r,c}/\Omega_{w,c}},\label{iner2}
\el
where the left hand side of \eqref{iner2} depends only on the ratio $(\eta_s/A_\text{eq})$, a parameter from the cRW patch, while the right hand side 
depends on the parameters of the quantum bouncing background. The small $w$ expansion is valid
when 
\bl
\frac{\left(6(\bar{a}_{\text{Form}}/\bar{a}_c)-(\Omega_{r,c}/\Omega_{w,c})
\right)^2}{96\left((\Omega_{r,c}/\Omega_{w,c})^2+6(\Omega_{r,c}/\Omega_{w,c})\right)}\gg w,\label{dewjioo}
\el
which ensures that the O($\omega^0$) terms in Eq.\eqref{kJcomi} dominate over the O($\omega$) terms.
The formation times of PBHs with masses $(10^{13}, 10^{2}, 10^{-14}, 10^{-17}) M_\odot$ satisfy condition  \eqref{dewjioo}, so Eq. \eqref{iner2} applies in these cases. 
When the following condition holds  
\bl
\bar{a}_{\text{Form}}/\bar{a}_c\gg \Omega_{r,c}/\Omega_{w,c},\label{equation78}
\el
Eq.\eqref{iner2} can be simplified to 
\bl
F\left(\frac{\eta_s}{A_\text{eq}}\right)^2G\left(\frac{\eta_s}{A_\text{eq}}\right)\approx w\frac{\pi^2}{24}.\label{iner3}
\el
The condition \eqref{equation78} implies that when the formation time is significantly displaced the 
dust-radiation equality, the lower bound on the surface coordinate $\chi_a$ and the critical curvature perturbation $\zeta_c$ depend only on the equation of state $w$, as seen in Table \ref{table1} for 
PBHs with masses $M>10^{-14} M_{\odot}$. For a PBH of mass $10^{-17} M_\odot$, the formation time is close to the 
dust-radiation equality, thus  $\chi_a$ and $\zeta_c$ 
differ from those of the other three masses. Plugging the value of \eqref{dewjio} and $w=9.44\times 10^{-22}$ into both sides of Eq. \eqref{iner3} shows the equation is satisfied.

Using \eqref{massfraction}, the mass fraction $\beta$ for a PBH of mass $10^{13} M_\odot$ is 
\bl
\beta (10^{13} M_\ensuremath{\odot} ) &
= \text{Erfc}[6.532\times 10^{54}]\approx0,\label{beta113}
\el
where the the curvature perturbation spectrum from Table \ref{table1N}, ${\cal P}_{\zeta}=2.826 \times 10^{-152}$, and the corresponding
critical curvature perturbation from Table \ref{table1},
$\zeta_c=1.553\times10^{-21}$ have been used. Applying the same procedure for other masses gives
\bl
\beta (10^{2} M_\ensuremath{\odot} ) &
= \text{Erfc}[1.419\times10^{42}]\approx0,\label{beta2}
\\
\beta (10^{-14} M_\ensuremath{\odot} ) &
= \text{Erfc}[3.060\times 10^{23}]\approx0,\label{betam14}
\\
\beta (10^{-17} M_\ensuremath{\odot} ) &
= \text{Erfc}[1.200\times 10^{20}]\approx0.\label{betam21}
\el
The mass fraction is negligible across a wide range of PBH masses, indicating that no significant PBH formation occurs before transition time during the contracting phase. Although all 
$\beta$ values in Eqs. \eqref{beta2}–\eqref{betam21} are extremely small, smaller PBHs (with lower masses and smaller physical scales) exhibit relatively higher formation rates. This follows because the argument of the complementary error function decreases as the mass and associated scale decrease. The mass fractions in Eqs. \eqref{beta113} $\sim$ \eqref{betam21} differ from the brown curve in Fig. 7 of our previous work \cite{Barroso2025} for the pure-dust quantum bouncing model, where mass fraction reaches unity for $M\lesssim10^{-12} M_{\odot}$. The difference arises from two reasons:  (i) The collapse process in the three-zone model differs from that based on the LTB solution, and the wave-propagation criteria used here yield a more restrictive threshold for PBH formation. (ii) Radiation pressure slows collapse and suppresses the amplitude of perturbation spectrum, thereby reducing the mass fractions.

\subsection{Mass fraction from Criterion 2}
\label{sub62}

In Sect. \ref{subsec5p2}, we discussed PBH formation under Criterion 2 and showed that the lower bound on the surface coordinate $\chi_a$ approaches that from the Criterion 1 as $\eta_s/A_\text{eq}\rightarrow0$ (Fig.\ref{dfrefgerThorZion}). Therefore, one  expects that the 
critical curvature perturbation $\zeta_c$ and the corresponding mass fraction $\beta$ obtained under Criterion 2 coincide with those from Criterion 1. 

In Sect. \ref{sub62}, we solve the coupled equations \eqref{coupledrquation} under the Criterion 2. Unlike Criterion 1, there is no analytic solution for $\chi_a$ in Eq.\eqref{frejiof}. Therefore, we solve the following equation numerically to obtain an interpolation for $\chi_a$ as a function of $\eta_s/A_\text{eq}$ 
{\footnotesize\bl
&\arcsin\left\{\left[1+\frac{\Big(\frac{\eta_s}{A_\text{eq}}\cos(\frac{\eta}{\eta_\text{eq}}\frac{\eta_s}{A_\text{eq}})+\sin(\frac{\eta}{\eta_\text{eq}}\frac{\eta_s}{A_\text{eq}})\Big)^2}{\Big(1-\cos(\frac{\eta}{\eta_\text{eq}}\frac{\eta_s}{A_\text{eq}})+\frac{\eta_s}{A_\text{eq}}\sin(\frac{\eta}{\eta_\text{eq}}\frac{\eta_s}{A_\text{eq}})\Big)^2}\right]^{-1/2}\right\}\nn
\\
&~~~~~~~~~~~~~~~~~~~~~-\frac{2\eta_{s}/A_\text{eq}}{\sqrt{\frac{9}{2}+3\frac{\eta_{s}^2}{A_\text{eq}^2}
+\frac{9}{2}\sqrt{1+\frac{\eta_{s}^2}{A_\text{eq}^2}}}}F\left[\frac12\left(\frac{\eta}{\eta_\text{eq}}\frac{\eta_s}{A_\text{eq}}-\frac{\eta_\text{max}}{L}\right) \Bigg|\frac{\sqrt{1+\eta_{s}^2/A_\text{eq}^2}}{\frac{1}{2}+\frac{\eta_{s}^2/A_\text{eq}^2}{3}
+\frac{1}{2}\sqrt{1+\eta_{s}^2/A_\text{eq}^2}}\right]=0,\label{halfcir}
\el}
where the first and second terms on the left-hand side represent the 
apparent horizon \eqref{dhewuihfdui} and the sound wave front \eqref{newfhuirewfheuirnew1}, respectively.
We first solve Eq. \eqref{halfcir} numerically to obtain the crossing time $(\eta/\eta_\text{eq})$ at which the sound and apparent horizons coincide. Substituting this value into  
the first term of \eqref{halfcir} gives the surface coordinate 
\bl
\chi_a=\chi_a \left(\frac{\eta_s}{A_\text{eq}}\right).\label{dhewuin}
\el
Plugging Eq.\eqref{dhewuin} 
into the right-hand side of Eq.\eqref{frejiof} 
and evaluating the left-hand side at the PBH formation time yields 
\bl
\frac14 R_{J}(\bar{a}_{\text{Form}})\bar{a}_{\text{Form}}=
\frac{2A_\text{eq}}{\eta_{s}^2/A_\text{eq}^2}\left(1+\sqrt{1+\eta_{s}^2/A_\text{eq}^2}\right)\sin\chi_{a}(\eta_s/A_\text{eq}).\label{frejiofn}
\el
\begin{table}[h]  
\centering  
\begin{tabular}{|c|c|c|c|c|c|}  
\hline  
\text{{\bf Criterion 2}}& $A_\text{eq} (s)$ & $\eta_s (s)$ & $\chi_a$ & $\zeta_c$ & $\beta$
\\ \hline  
$10^{13} M_{\ensuremath{\odot}}$ 
 &   $1.105 \times10^{15}$ & 
 $5.452\times 10^{3}$  & 
 $7.880 \times 10^{-11}$  & $1.552\times 10^{-21}$ & 0
 \\ \hline  
$10^{2} M_{\ensuremath{\odot}}$ & $1.106\times10^{4}$ & 
$5.459\times 10^{-8}$  &   
$7.882\times 10^{-11}$ & 
$1.553 \times 10^{-21}$ & 0
  \\ \hline  
$10^{-14} M_{\ensuremath{\odot}}$  
& $1.104\times10^{-12}$  & 
$5.457\times 10^{-24}$ &  
$7.894\times 10^{-11}$  & 
$1.558\times 10^{-21}$ & 0
\\
\hline 
$10^{-17} M_{\ensuremath{\odot}}$ 
& $1.087 \times 10^{-15}$  
& $6.329 \times10^{-27}$ 
& $9.237\times 10^{-11}$ 
& $2.133\times 10^{-21}$ & 0
\\
\hline 
\end{tabular}  
\caption{Values of $(A_\text{eq}, \eta_s, \chi_a, \zeta_c,\beta)$ based on  criterion 2 for PBHs of different masses. }\label{table2}
\end{table}
Finally, using Eq. \eqref{frejiofn} together with Eq. \eqref{dewhduiwe22}, and following the same procedure described below Eq. \eqref{dewhduiwe22}, we numerically derive the quantities  $(A_\text{eq}, \eta_s, \chi_a, \zeta_c,\beta)$. The results are listed in Table \ref{table2}.

We find that the ratio $\eta_s/A_\text{eq}$ is the same for masses $10^{13}, 10^2, 10^{-14} M_{\ensuremath{\odot}}$ 
\bl
\frac{\eta_s}{A_\text{eq}}\approx 4.943 \times 10^{-12},
\el
which implies that the critical curvature perturbation is identical for these three masses. The mass fractions for four representative masses 
are shown in the last column of Table \ref{table2}, they vanish  and match those from Criterion 1. This agreement confirms the internal consistency between the two criteria.  Note that neither result depends on the bounce scale factor $\bar{a}_\textsc{B}$. The PBH formation found here does not rely on the bouncing mechanism.

\section{Conclusions and Discussion}\label{sec7}

In this paper we developed a framework to study primordial black hole formation during the contracting phase of spatially flat dust--radiation bouncing cosmologies. 
Although we employ the canonical quantum bouncing mechanism, the results are independent of bounce details because PBH formation is studied in the classical contracting phase. For a representative mass spectrum $(10^{13}, 10^{2}, 10^{-14}, 10^{-17}) M_{\odot}$, we applied the Press--Schechter formalism to estimate the PBH mass fraction $\beta$ at formation. Our results show that the mass fractions approach zero, indicating a negligible probability of PBH formation before the transition time $\bar a_{\rm Tr}$, which occurs shortly after dust--radiation equality along the contracting evolution. 
Quantitatively, we obtain a critical curvature perturbation  $\zeta_c \approx (1.55 \sim 2.13)\times 10^{-21}$ and an extremely suppressed curvature spectrum at formation, $P_\zeta(k_J,\bar a_{\rm Form}/\bar a_c)\approx 10^{-152} \sim 10^{-82}$, which yields $\beta(M)=\mathrm{Erfc}\!\left[\zeta_c/\sqrt{2P_\zeta}\right]\approx 0$ with arguments $10^{20} \sim 10^{54}$. 
These results differ from those in the pure-dust quantum bouncing model in our previous work \cite{Barroso2025} (see the brown curve in Fig.~\ref{ccfrefgerThorZion}), where the mass fraction approaches unity for $M \lesssim 10^{-12} M_{\odot}$ and vanishes for $M \gtrsim 10^{-9} M_{\odot}$. The discrepancy arises because the collapse model and sound wave propagation criteria in this work raise the PBH formation threshold, and radiation pressure further slows collapse and suppresses the amplitude of the curvature perturbation spectrum at formation time compared to the pure-dust case. Beyond the numerical estimates of $\beta$, the formalism we developed here, treating PBH formation in a quantum bouncing background with dust and radiation, is nontrivial and readily generalizable to other cosmological settings and multi-component fluids. We summarize and conclude as follows.

The curvature perturbation spectrum was obtained by numerically computing the scale factor and applying an adiabatic semi-analytical method that expresses the spectrum in terms of the first and second time derivatives of the scale factor. This approach avoids numerical instabilities arising from the extremely small equation-of-state parameter $w = 9.440\times 10^{-22}$. We found that the numerical accuracy of the spectrum at PBH formation decreases with increasing PBH mass, since the adiabatic approximation degrades on larger scales corresponding to more massive PBHs. The spectrum amplitude at formation also decreases with PBH mass for two reasons. First, the leading-order spectrum is proportional to $k^3$, so smaller PBHs (larger $k$) have larger amplitudes. Second, more massive PBHs form earlier in the contracting phase, when the scale factor is larger, while the spectrum decreases as $\bar{a}$ departs from the bounce point $\bar{a}_\textsc{B}$.

Different PBH sizes correspond to different curvature perturbation spectra at formation time. The size is set by the Jeans scale, which defines the minimum perturbation length capable of collapsing under gravitational instability. We generalized the dynamical-system analysis of gravitational instability for two non-relativistic fluids presented in Refs.~\cite{2jeans_static,2jeans_cosmo,2jeans_short}. The radiation component is relativistic, so the starting equations in those works are not directly applicable. The Jeans wavelength derived here is shorter than the Hubble radius before the transition time and longer afterwards. Around the transition time, the Jeans wavelength shows a sharp but continuous transition between the dust and radiation limits. After the transition, the Jeans length exceeds the Hubble radius.

We generalized the three-zone model, developed for a single fluid \cite{Harda2013}, to describe the evolution of an overdense region in a two-fluid quantum bouncing cosmology. The overdense region is modeled as a closed Robertson--Walker spacetime, and the model contains two free parameters, $\eta_s$ and $A_\text{eq}$. We derived the critical curvature perturbation, $\zeta_c$, in the three-zone model using two criteria. Criterion~1 compares the sound propagation timescale to the time of maximum expansion of the cRW patch. Criterion~2 compares the sound propagation time to the time of apparent horizon formation. We express $\zeta_c$ as a function of $\eta_s$ and $A_\text{eq}$. In general, Criterion~1 yields the same $\zeta_c$ for both expanding and contracting backgrounds. Under Criterion~2, however, $\zeta_c$ in a contracting background is smaller than in an expanding background, indicating that PBHs form more easily in the contracting phase, as expected. The two criteria yield identical values of $\zeta_c$ in the limit $\eta_s/A_\text{eq}\rightarrow 0$.

In general, by matching the scale factor and energy density at the time of maximum expansion of the cRW patch to those of the quantum bouncing model at the time of PBH formation, with a one-to-one correspondence set by the Jeans scale for a given PBH mass, we expressed the parameters $\eta_s$ and $A_\text{eq}$ in terms of the quantities of the quantum bouncing model. Using these parameters we derived the critical curvature perturbation $\zeta_c(\eta_s, A_\text{eq})$ under two criteria. For the parameters considered, both criteria yield extremely small and consistent results. Since $\zeta_c$ is much larger than the square root of the variance of the curvature perturbation (the power spectrum), only a negligible fraction of perturbations can collapse to form PBHs. Consequently, PBH formation via gravitational instability is essentially negligible before the transition time. After this time, the Jeans length exceeds the Hubble radius, so perturbations typically become super-Hubble before becoming super-Jeans; as a result, gravitational collapse on sub-Hubble scales is prevented and PBH formation by gravitational instability is again disfavored. Within the assumptions adopted here (adiabatic-dominated perturbations and Gaussian statistics in the Press--Schechter estimate), an alternative mechanism would therefore be required to amplify the curvature spectrum sufficiently to produce PBHs, and we leave the exploration of such mechanisms to future work.

\

\textbf{Acknowledgements}

\

We thank Yi-Fu Cai and Mian Zhu for their valuable discussion and support. Xuan Ye is supported in part by NSFC Grant No. 12433002. Luiz Felipe Demétrio acknowledges the support of CAPES under the grant DS 88887.902808/2023-00. 
Eduardo José Barroso acknowledges support from the Enigmass+ research federation (CNRS, Université Grenoble Alpes, Université Savoie Mont-Blanc).  This work is supported by the French National Research Agency in the framework of the “France 2030” (ANR-15-IDEX-02).
Sheng-Feng Yan acknowledges the support of the postdoctoral fellowship program of CPSF under grant number GZC20240212.
Nelosn Pinto-Neto acknowledges the support of CNPq of Brazil under grant PQ-IB 310121/2021-3.

\

\bibliographystyle{unsrt}
\bibliography{g2.bib}

\appendix

\numberwithin{equation}{section}

\section{ Adiabatic Curvature Perturbation Power Spectrum }\label{wkbaPP}

In this Appendix we derive the effective frequencies and the adiabatic power  spectrum of the curvature perturbation up to second adiabatic order, which are used in Sect. \ref{sec22}.

Assume the adiabatic mode function $u_k$ takes the
form 
\bl
u_k =\frac{ C_k}{\sqrt{2 \Omega_{k}(x)}} e^{-i \int^{x}\Omega_{k}(x') dx'},\label{dew}
\el
where $\Omega_{k}(x)$ is the dimensionless effective frequency and 
$C_k$ is a coefficient with dimension of the square root of time in natural units, to be determined 
later by comparison with \eqref{dhewuiuh28}. Plugging \eqref{dew} in Eq.\eqref{ULH} yields 
\bl
\Omega_{k}=\sqrt{\nu_{\zeta n}^2-\frac12\frac{\ddot{m}_{\zeta n}}{m_{\zeta n}}+
\frac14 \left(\frac{\dot{m}_{\zeta n}}{m_{\zeta n}}\right)^2-
\frac{ \ddot{\Omega}_{k}}{2\Omega_{k}}
+\frac{3 \dot{\Omega}_{k}^2}{4\Omega_{k}^2}
},\label{equfhiu}
\el
where $\nu_{\zeta n}$ and $m_{\zeta n}$ are defined in Eqs.\eqref{effecitfreqn} and \eqref{mzetatfreqn}, respectively.
The number of time derivatives indicates the adiabatic order \cite{Chakraborty,Parker_Toms_2009, Zhang2020}. Since 
$\nu_{\zeta n}$ and $m_{\zeta n}$ involve no time derivatives, they are 
of 0th order. Basing on this, Eq.\eqref{equfhiu} can be solved iteratively.

We expand the effective frequency as 
\bl
\Omega_{k}=\Omega_{k}^{(0)}
+\Omega_{k}^{(2)}
+...,\label{eqw}
\el
where the superscript $(n)$ denotes the adiabatic order. 
Plugging \eqref{eqw} into \eqref{equfhiu} and keeping terms up to 2nd order yields 
the 0th-order effective frequency
\bl
\Omega_{k}^{(0)}=\nu_{\zeta n}.\label{0th}
\el
and the 2nd-order effective frequency 
\bl
\Omega_{k}^{(2)}&=-\frac1{4\nu_{\zeta n}}\frac{\ddot{m}_{\zeta n}}{m_{\zeta n}}+
\frac1{8\nu_{\zeta n}} \left(\frac{\dot{m}_{\zeta n}}{m_{\zeta n}}\right)^2-
\frac{ \ddot{\nu}_{\zeta n}}{4\nu_{\zeta n}^2}
+\frac{3 \dot{\nu}_{\zeta n}^2}{8\nu_{\zeta n}^3},\label{Omnewgkh2}
\el
where \eqref{0th} has been used. Next we determine the 
coefficient $C_k$. In the far past, $\bar{\eta}\rightarrow -\infty$, 
the 0th order adiabatic mode function becomes 
\bl
\lim_{\bar{\eta}\rightarrow -\infty}u_k =\frac{ C_k}{\sqrt{2\sqrt{w} k_H }} e^{-i \sqrt{w} k_H x },\label{dewdew}
\el
where \eqref{0th} and Eq. \eqref{effecitfreqn} have been used. Comparing 
Eq. \eqref{dewdew} with 
Eq. \eqref{dhewuiuh28} yields 
\bl
C_k=\frac{1}{\sqrt{2\bar{a}_cH_0}}.\label{deterck}
\el
The dimensionless adiabatic spectrum of the curvature perturbation is  
\bl
{\cal P}_{\zeta }&=\frac{k^3}{2\pi^2}\frac{|u_{k}|^2}{m_{\zeta}}
\approx\frac{k^3}{8 \pi^2 \bar{a}_cH_0 m_{\zeta}}\frac{1}{|\Omega_{k}|},\label{defination1}
\el
where \eqref{dew} and \eqref{deterck} have been used.
Plugging \eqref{eqw} into \eqref{defination1} and expanding to the 2nd order gives
\bl
{\cal P}_{\zeta}&\approx\frac{k^3}{8 \pi^2 \bar{a}_cH_0 m_{\zeta} \nu_{\zeta n}}\left(1-\Omega^{(2)}_{k}/\nu_{\zeta n}\right),\label{denuw}
\el
where the adiabatic approximation condition 
\bl
\Omega^{(2)}_{k}/\Omega^{(0)}_{k}\ll 1,\label{WKBcodition}
\el
has been used.
Inserting \eqref{Omnewgkh2} into 
\eqref{denuw} and using the rescaled quantities $k_H\equiv k/(\bar{a}_c H_0 )$ and $m_{\zeta n}\equiv m_{\zeta } G/\bar{a}_c^2$
yields the curvature perturbation spectrum \eqref{dewhiutext}.

\section{Jeans Instability in a General Relativistic Mixture of Two Fluids}\label{B}

In this Appendix, we derive the equations governing the density contrasts of multi‑component fluids in General Relativity and examine the validity of the WKB approximation employed in the calculation of the Jeans wavenumber.

\subsection{Relativistic equation of density contrasts for multiple fluids }

The Jeans instability for two fluids in the static Newtonian limit was first studied using a dynamical‑system approach in Ref.~\cite{2jeans_static}. These results were later generalized to an expanding spacetime in Refs.~\cite{2jeans_cosmo,2jeans_short}. However, the equations for the density contrasts used in Refs.~\cite{2jeans_cosmo,2jeans_short} are valid only for non‑relativistic fluids with a small equation of state, $\omega_\lambda = \bar{p}_\lambda / \bar{\rho}_\lambda \approx 0$. Since our model includes radiation, a relativistic fluid, it is necessary to extend the results of Refs.~\cite{2jeans_cosmo,2jeans_short} to the relativistic case.

We start from the equations for the density contrasts of multiple fluids $\delta_{\lambda k} \equiv {\delta \rho_{\lambda k}}/{\bar\rho}$, with $\lambda = 1, 2,...$ in General Relativity. Following Ref. \cite{maggiore2018gravitational}, the Einstein equations for scalar perturbations in an $N$-fluid system within a perturbed fRW spacetime (Newtonian gauge)
read
\begin{subequations}
    \begin{align}
        k^2 {\Phi}_k+3 \bar{\mathcal{H}}\left({\Phi}_k^{\prime}+\bar{\mathcal{H}} {\Phi}_k\right) & =4 \pi G \bar{a}^2   \sum_{_{\sigma}=1}^{N} {\bar{\rho}}_{\sigma} {\delta}_{\sigma k} \, , \label{pertA} \\ 
        {\Phi}_k^{\prime}+\bar{\mathcal{H}} {\Phi}_k & =-\frac{4 \pi G \bar{a}^2}{k^2} \sum_{_{\sigma}=1}^{N}{\bar{\rho}}_{\sigma}\left(1+w_{\sigma}\right) v_{\sigma k} \, , \label{pertB} \\ 
        {\Phi}_k^{\prime \prime}+3 \bar{\mathcal{H}} \tilde{\Phi}_k^{\prime}+\left(\bar{\mathcal{H}}^2+2 \bar{\mathcal{H}}^{\prime}\right) {\Phi}_k & =-4 \pi G a^2 \sum_{{\sigma}=1}^{N} w_{\sigma} {\bar{\rho}}_{\sigma} v_{\sigma k} \, \label{pertC},
    \end{align}
\end{subequations}
where $\Phi_k$ is the Bardeen gauge invariant potential and $v_{\sigma k}$ is the scalar velocity perturbation associated with fluid $\sigma$. The perturbed continuity and Euler equations for each fluid are \cite{maggiore2018gravitational}
\begin{subequations}
    \begin{align}
        {\delta}_{\lambda k}^{\prime} & = -\left(1+w_\lambda\right)\left({v}_{\lambda k}+3 {\Phi}_k^{\prime}\right) \label{pertD} \, ,  \\
        {v}_{\lambda k}^{\prime}+ \bar{\mathcal{H}}\left(1-3 w_\lambda\right)v_{\lambda k}& =k^2\frac{w_\lambda}{1+w_\lambda} {\delta}_{\lambda k}-k^2{\Phi}_k.\label{pertE}
    \end{align}
\end{subequations}
Differentiating \eqref{pertD} with respect to time, eliminating $\Phi_k$ using \eqref{pertA}–\eqref{pertC}, and eliminating $v_{\lambda k}$ using \eqref{pertE}, one obtains, after straightforward but lengthy algebra, the fully relativistic evolution equations for the density contrasts of each fluid component,
\begin{align}
    \delta^{\prime\prime}_{\lambda k} + & (1-3w_{\lambda})\bar{\mathcal{H}}\delta^{\prime}_{\lambda k} + w_{\lambda}k^{2}\delta_{\lambda k} = 4\pi G\bar{a}^{2}\Bigg\{(1+w_{\lambda})\Big[ (2+3w_{\lambda})\sum_{\sigma = 1}^{N}\bar{\rho}_{\sigma}\delta_{\sigma k} + 3\sum_{\sigma = 1}^{N}w_{\sigma}\bar{\rho}_{\sigma}\delta_{\sigma k} \Big]+ \nonumber 
    \\ 
    & \quad + 3\frac{(1+w_\lambda)\left((1+3w_{\lambda})(k^{2}+3\bar{\mathcal{H}}^{2}) - 6\bar{\mathcal{H}}^{\prime}\right)}{ 3\dpar{{\bar{\mathcal{H}}^{\prime} - \bar{\mathcal{H}}^{2} } }\dpar{k^{2} + 3\bar{\mathcal{H}}^{2}} + k^{4} }\left[ \left(\bar{\mathcal{H}}^{2} - \bar{\mathcal{H}}^{\prime} - \frac{k^{2}}{3}\right)\sum_{\sigma = 1}^{N}\bar{\rho}_{\sigma}\delta_{\sigma k} + \bar{\mathcal{H}}\sum_{\sigma = 1}^{N}\bar{\rho}_{\sigma}\delta^{\prime}_{\sigma k} \right] \Bigg\} \label{contD} \, ,
\end{align}
where $\lambda = 1,2, ...,N$ is a free index referring to each fluid.  The coupled system  is rather involved. Since the Jeans instability is relevant on sub‑Hubble scales, we adopt the short‑wavelength approximation: $k/\bar{\cal H} \gg 1, \bar{\cal H}^{\prime}/\bar{\cal H}^{2} \ll 1 $. In this regime, \eqref{contD} reduces to the simpler form
\begin{equation}
    \delta^{\prime\prime}_{\lambda k} + (1-3w_{\lambda})\bar{\cal H}\delta^{\prime}_{\lambda k} + w_{\lambda}k^{2}\delta_{\lambda k} = 4\pi G\bar{a}^{2}(1+w_{\lambda}){ \sum_{\sigma = 1}^{N}{(1 + 3w_{\sigma}  ){\bar{\rho}_\sigma\delta_{\sigma k}}} }  \, .\label{contDsubH}
\end{equation}
which reduces the forms used in Refs. \cite{2jeans_cosmo,2jeans_static} in the non-relativistic limit $w_{\lambda}\rightarrow 0$. Therefore, Eq. \eqref{contDsubH} is a natural generalization of the two‑fluid Jeans equations that includes a relativistic component.

Introducing the rescaled variables
$\tilde{\delta}_{\lambda}
=\bar{a}^{\frac{1-3\omega_\lambda}{2}}\delta_\lambda$ into 
\eqref{contDsubH} yields 
\bl
\tilde{\delta}_\lambda''+\left[\omega_\lambda k^2 +
\frac{2\pi G}{3}\bar{a}^2(1-3\omega_\lambda)\sum_{\sigma=1}^{N}\bar{\rho}_\sigma(\omega_{\lambda}-3\omega_\sigma-1)\right]
\tilde{\delta}_{\lambda k}
=4\pi G (1+\omega_\lambda)
\bar{a}^{\frac{(4-3\omega_\lambda)}{2}}\sum_{\sigma=1}^{N}
(1+3\omega_\sigma)\bar{\rho}_\sigma\bar{a}^{\frac{3\omega_\sigma}{2}}
\tilde{\delta}_{\sigma k},\label{B5}
\el
where we have used the first and second Friedmann equations for a fRW spacetime.
For the dust–radiation system with 
$N=2$, labeling the components by $\lambda=\omega$ (dust, $w_\omega=w$) and $\lambda=r$ (radiation, $w_r=1/3$), 
 Eq.~\eqref{B5} reduces to Eqs.~\eqref{app_dynamical_system1} and \eqref{app_dynamical_system2}.

\subsection{Validity of the WKB approximation}\label{subb2}

In Sect.~\ref{sec3}, we used the WKB approximation to derive the first‑order solution of the Jeans wavenumber, Eq.~\eqref{jeanss2t}. In what follows, we examine the validity of this approximation. To this end, we introduce the transformation matrix $\textbf{A}\equiv (\bm{ \xi}_1,\bm{ \xi}_2,\bm{ \xi}_3,\bm{ \xi}_4)$, where  $\bm{\xi}_{j}$ is the eigenvector of matrix  $\bm{T}$ defined in Eq. \eqref{MartixT}.
Each eigenvector takes the form $(\lambda_j\beta_j, \beta_j, \lambda_j,1)^\text{T} $ with eigenvalues $\lambda_j$ given in Eqs.~\eqref{eigv1t} and \eqref{eigv3t}, and coefficients $\beta_j$ defined by
\begin{subequations}
    \begin{align}
        \beta_1 = \beta_2& = -\frac{1}{{2}}\dpar{h(k,\bar{a}) - \sqrt{h(k,a)^2+4z(k,a)}}\label{eigv1} , \\
        \beta_3 = \beta_4& = -\frac{1}{{2}}\big(h(k,\bar{a}) + \sqrt{h(k,\bar{a})^2+4z(k,\bar{a})}\big)\label{eigv3},
    \end{align}
\end{subequations}
with $h(k,\bar{a})$ and $z(k,\bar{a})$ being given by
\begin{subequations}
    \begin{align}
        h(k,\bar{a}) & \equiv \frac{1}{16\pi G\bar{a}^{3/2}\bar{\rho}_w}\dcol{\dpar{1-3w} k^{2} + 2\pi  G\bar{a}^2\dpar{6\bar{\rho}_w - 17\bar{\rho}_r} }, \\ 
        z(k,\bar{a}) & \equiv \bar{a}\frac{3\bar{\rho}_r}{2\bar{\rho}_w} \, .
    \end{align}
\end{subequations}
One can write down the 
explicit expression of the  matrix equation \eqref{fjheurih} 
\begin{equation}\label{maticequation}
\frac{d}{d\bar{\eta}}\begin{pmatrix}
    y_1 \\
    y_2  \\
    y_3  \\
    y_4 
  \end{pmatrix}
  =\left[\begin{pmatrix}
    \lambda_1 & 0 & 0 & 0 \\
    0 & -\lambda_1 & 0 & 0 \\
    0 & 0 & \lambda_3 & 0 \\
    0 & 0 & 0 & -\lambda_3
  \end{pmatrix}
  -\begin{pmatrix}
    \epsilon_1 & \epsilon_2 & \epsilon_3 & \epsilon_4 \\
    \epsilon_2 & \epsilon_1 & \epsilon_4 & \epsilon_3 \\
    \epsilon_5 & \epsilon_6 & \epsilon_7 & \epsilon_8 \\
        \epsilon_6 & \epsilon_5 & \epsilon_8 & \epsilon_7
  \end{pmatrix}\right]\begin{pmatrix}
    y_1 \\
    y_2  \\
    y_3  \\
    y_4 
  \end{pmatrix},    
\end{equation}   
where the $\epsilon_j$ coefficients measure the corrections to the diagonalization and are given by
    \begin{align}
        \epsilon_1 & \equiv\frac{\lambda_1'}{2\lambda_1}+\frac{\beta_1'}{\beta_{13}}, 
        ~~~\epsilon_2  \equiv-\frac{\lambda_1'}{2\lambda_1}, 
        ~~~\epsilon_3  \equiv\frac{\beta_3'}{2\beta_{13}}\left(1+\frac{\lambda_3}{\lambda_1}\right), 
        ~~~\epsilon_4  \equiv\frac{\beta_3'}{2\beta_{13}}\left(1-\frac{\lambda_3}{\lambda_1}\right), 
        \\
        \epsilon_5 & \equiv-\frac{\beta_1'}{2\beta_{13}}\left(1+\frac{\lambda_1}{\lambda_3}\right), 
        ~~~\epsilon_6  \equiv-\frac{\beta_1'}{2\beta_{13}}\left(1-\frac{\lambda_1}{\lambda_3}\right), 
       ~~~ \epsilon_7  \equiv\frac{\lambda_3'}{2\lambda_3}-\frac{\beta_3'}{\beta_{13}}, 
        ~~~\epsilon_8  \equiv-\frac{\lambda_3'}{2\lambda_3} \, . \label{dynamical1}
    \end{align}
The WKB solution for {\bf y}, given by \eqref{capitaly}, is valid when 
 \begin{align}
                \label{lambamoine1}
        \left| \epsilon_i / \lambda_j\right| & \ll 1 \, , i = 1, 2...8 \, \, \text{and} ~ j= 1, 3.
\end{align}
In what follows, we show that for the the scales of interest (see Table~\ref{table1N}), condition 
\eqref{lambamoine1} is satisfied when $k > k_J$ or $a > a_J$, provided the modes remain sub‑Hubble.
\begin{figure}[htb]
\center
 \subcaptionbox{}
   {%
     \includegraphics[width = .47\linewidth]{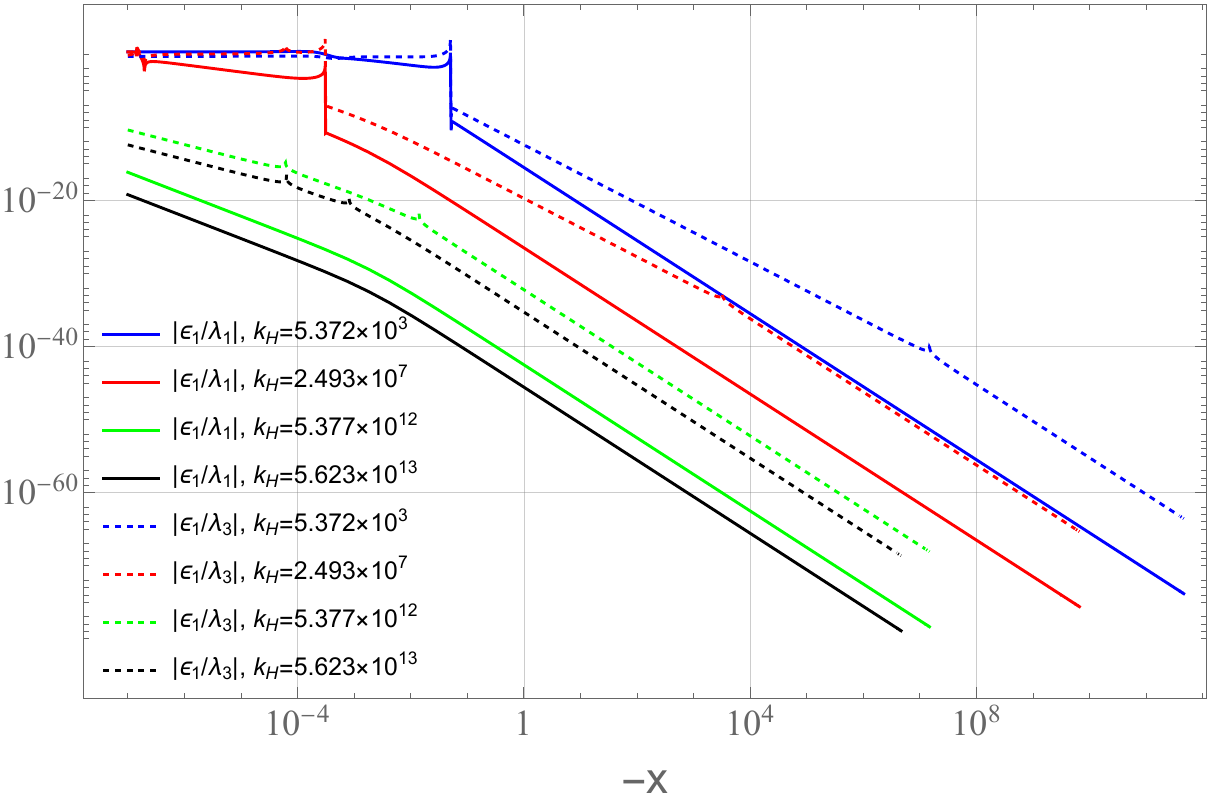}}
 \subcaptionbox{}
   {%
     \includegraphics[width = .47\linewidth]{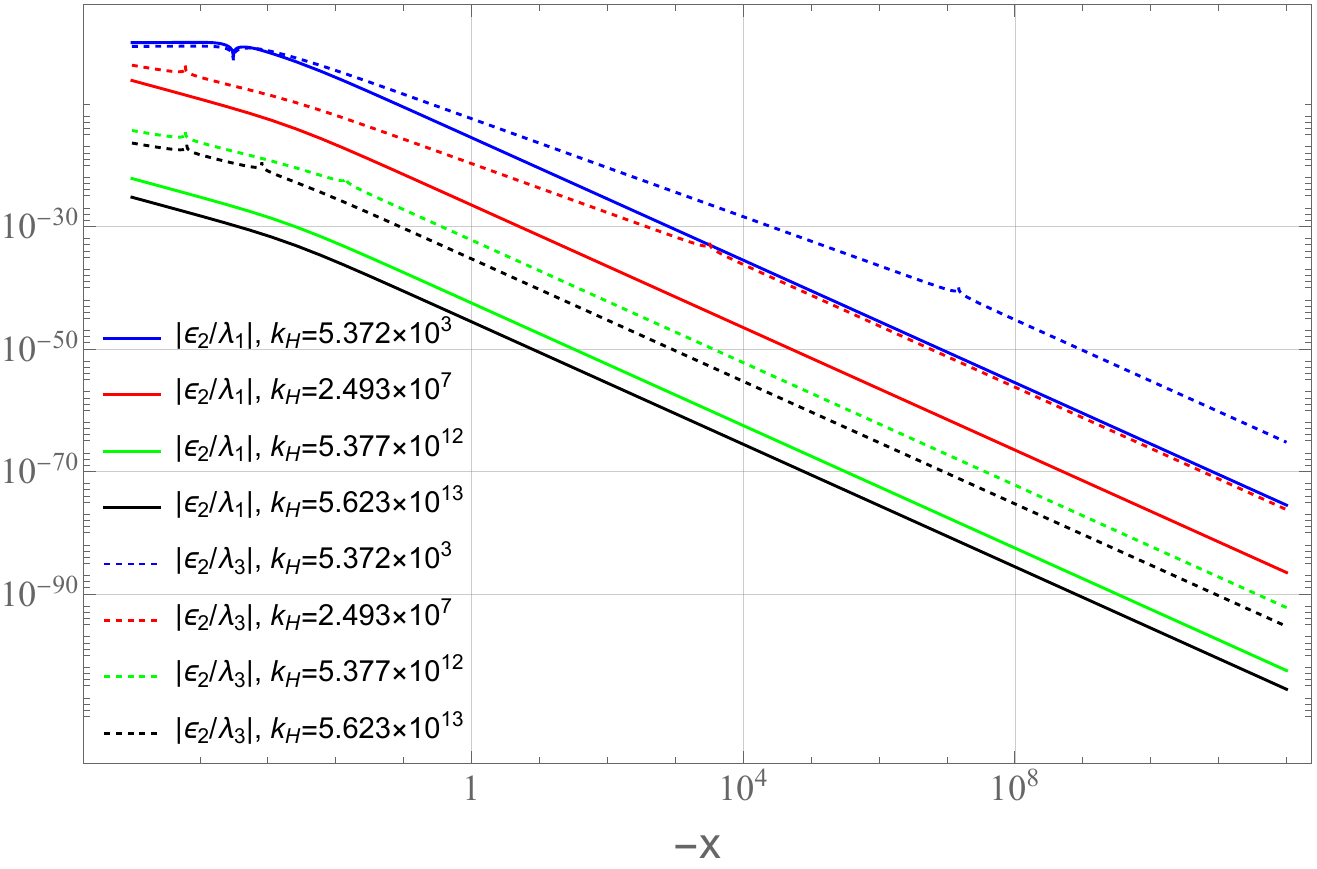}}
\subcaptionbox{}
   {%
     \includegraphics[width = .47
\linewidth]{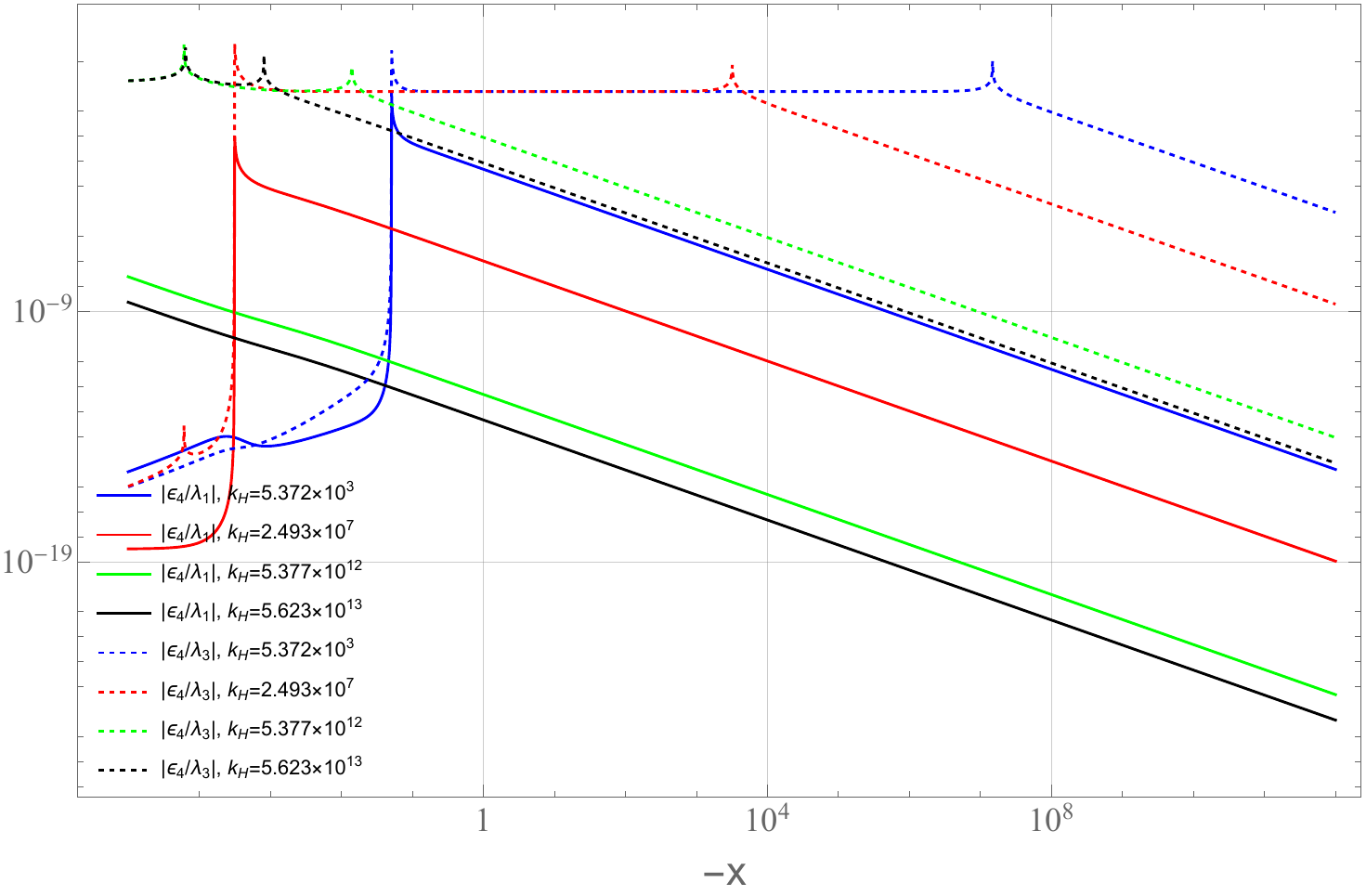}} 
 \subcaptionbox{}
   {%
     \includegraphics[width = .47\linewidth]{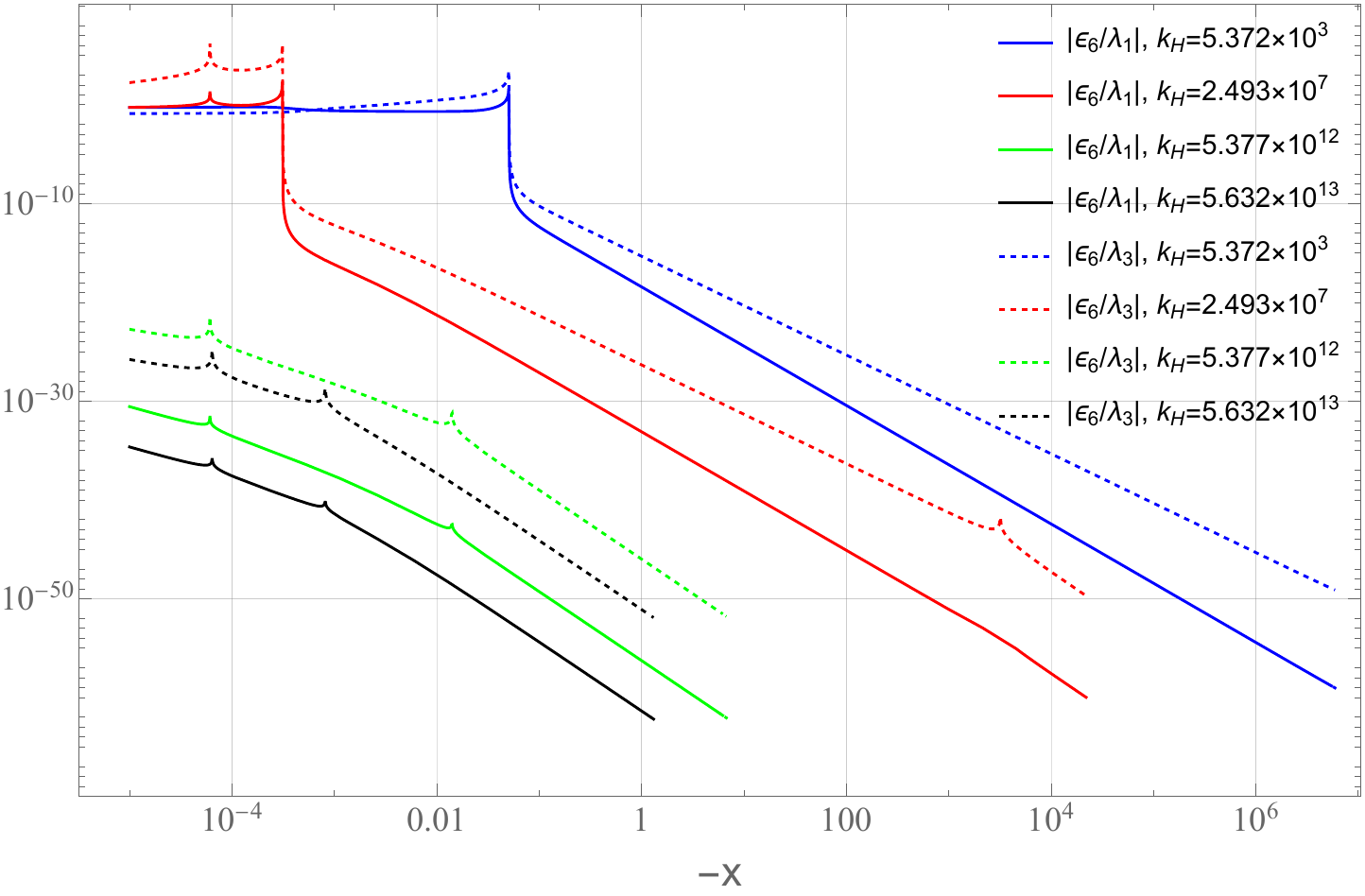}}
 \subcaptionbox{}
   {%
     \includegraphics[width = .47
\linewidth]{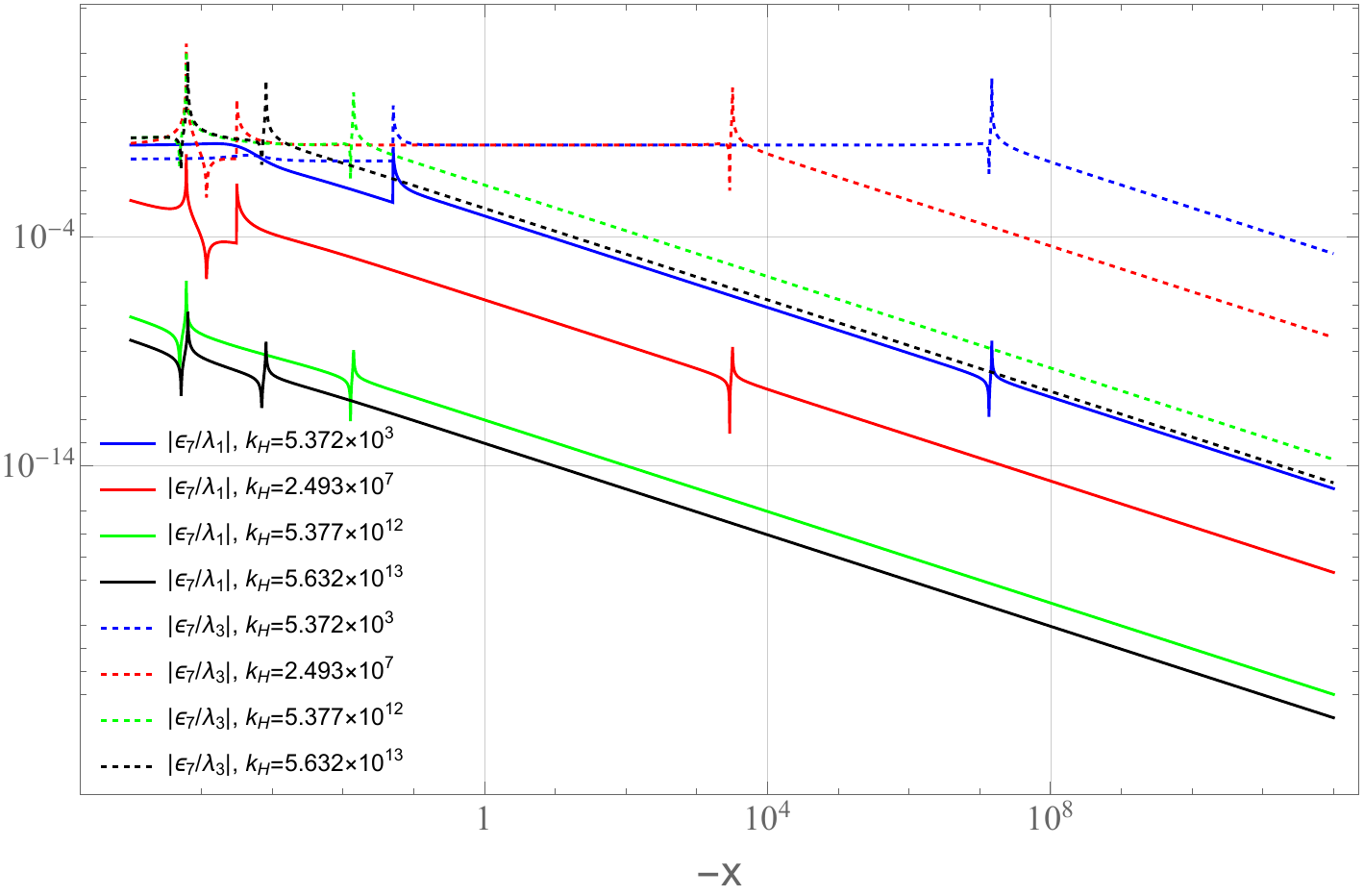}} 
\subcaptionbox{}
   {%
     \includegraphics[width = .47
\linewidth]{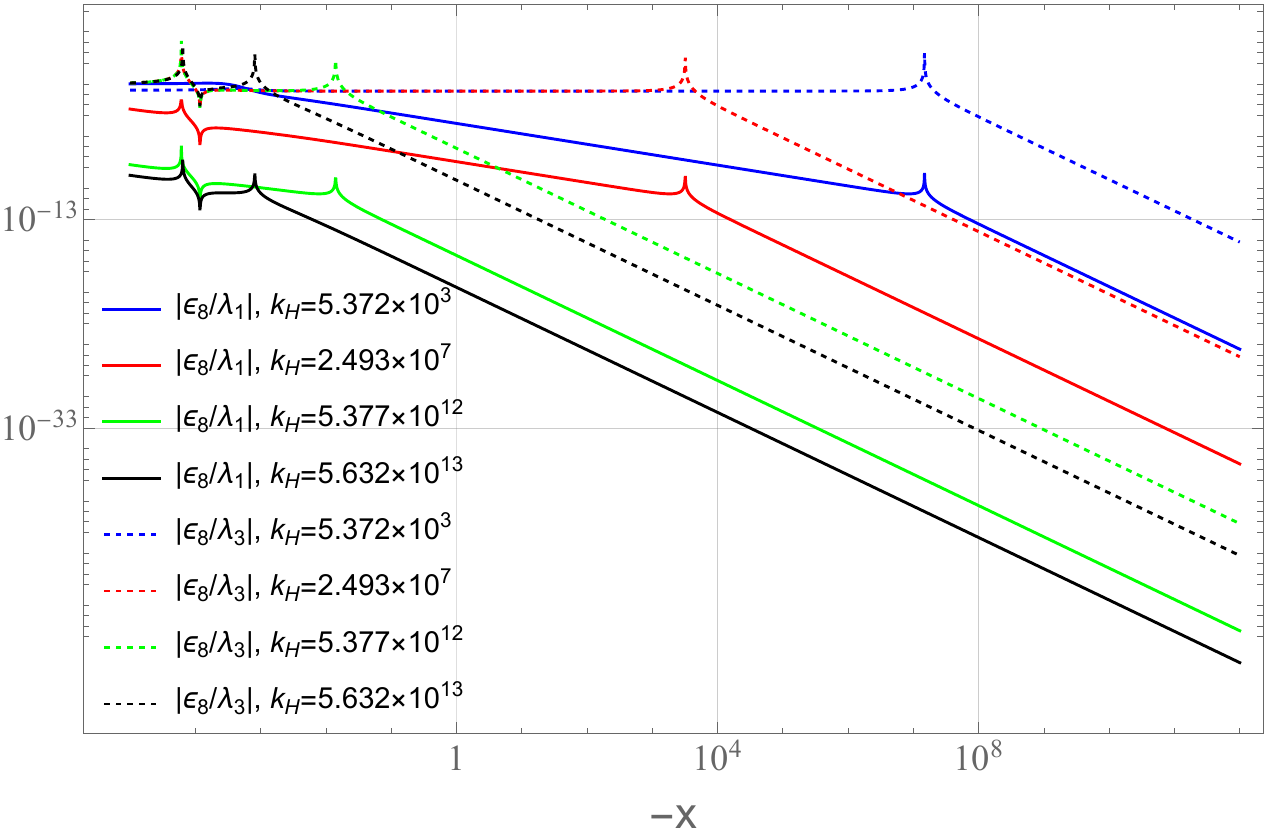}} 
\caption{(a) $\sim$ (f) Ratios 
 $|\epsilon_{1}/\lambda_{i}|$.
 $|\epsilon_{2}/\lambda_{i}|$.
 $|\epsilon_{4}/\lambda_{i}|$.
 $|\epsilon_{6}/\lambda_{i}|$.
 $|\epsilon_{7}/\lambda_{i}|$.
 $|\epsilon_{8}/\lambda_{i}|$, respectively.
Solid line corresponds to $i=1$, dashed line to $i=3$.
Blue: $k_H=5.372\times 10^3$. Red: $k_H=2.493\times 10^7$. Green: $k_H=5.377\times 10^{12}$. Black：$k_H=5.632\times 10^{13}$.
} \label{enumberL}
\end{figure}
In this regime, one can perform the stability analysis following Refs.~\cite{2jeans_static,2jeans_cosmo,2jeans_short}.
Rewriting the expressions for $\lambda_1$ and $\lambda_3$  from Eq.~\eqref{eigv3t} as
\begin{subequations}
    \begin{align}
        \lambda_{1,3} & = -\frac{1}{\sqrt{2}}\sqrt{f \mp \sqrt{f^2+4g}}\label{eigv1tn},
    \end{align}
\end{subequations}
where
\begin{subequations}
    \begin{align}
        f& =
\frac34(H_0\bar{a}_c)^2\left\{\left[2
        \bar{\Omega}_{w,c}\left(\frac{\bar{a}_c}{\bar{a}}\right)^{(1+3w)}
    +5\bar{\Omega}_{r,c}\left(\frac{\bar{a}_c}{\bar{a}}\right)^{2}
 \right]-\frac{4}{3}\dpar{w+1/3}k_H^2\right\}\, , \\ 
g&=\frac{(\bar{a}_cH_0)^4 }{3} \left\{18 \Omega_{w,c}\Omega_{r,c}\left(\frac{\bar{a}_c}{\bar{a}}\right)^{3(1+w)}-\left[k_H^2-12  \Omega_{r,c} \left(\frac{\bar{a}_c}{\bar{a}}\right)^2 \right] \left[k_H^2 w+\frac{1}{4}  \left(\Omega_{r,c}\left(\frac{\bar{a}_c}{\bar{a}}\right)^2 -6\Omega_{w,c}\left(\frac{\bar{a}_c}{\bar{a}}\right)^{1+3w}  \right)\right]\right\} \, ,
    \end{align}
\end{subequations}
where $\bar{\rho}_w$ and $\bar{\rho}_r$ are rescaled at time $\bar{a}_c$. Furthermore, $\beta_1$ and $\beta_3$ can be re‑written as
as 
\bl
\beta_{1,3}=-\frac{1}{{2}}\dpar{h \mp \sqrt{h^2+4z}},\label{c13}
\el
where 
\begin{subequations}
\bl
h&=\frac{ \bar{a}_c^{1/2}}{
6\Omega_{w,c}(\bar{a}_c/\bar{a})^{3/2+3w}}\left\{\dpar{1-3w} k_H^{2} +\frac34\Big[6\left(\frac{\bar{a}_c}{\bar{a}}\right)^{1+3w} - 17\left(\frac{\bar{a}_c}{\bar{a}}\right)^2\Big] \right\},\label{c14a}
\\
z&=\bar{a}_c\frac{3 \Omega_{r,c}}{2\Omega_{w,c}}\left(\frac{\bar{a}}{\bar{a}_c}\right)^{3w}.\label{c14b}
\el
\end{subequations}
To examine whether the condition \eqref{lambamoine1} is valid, 
we plot Fig.~\eqref{enumberL}(a)–(f) for illustration. 
Note that we do not plot $\epsilon_4$ and $\epsilon_6$, as they are of the same 
order of magnitude as $\epsilon_3$ and $\epsilon_5$, respectively.

The formation times of PBHs with $k_H=5.372\times 10^3$, $k_H= 2.493\times 10^7 $, $k_H= 5.377 \times 10^{12}$ and $k_H= 5.632 \times 10^{15}$ 
are $(-x)=1.484\times 10^{7}$, $(-x)=3.197 \times 10^3$, $(-x)=1.408\times 10^{-2}$ and $(-x)=8.159\times 10^{-4}$, respectively (see Table \ref{table1N}). 
For $|\epsilon_1/\lambda_j|$, $|\epsilon_2/\lambda_j|$, and $|\epsilon_6/\lambda_j|$ and for the small‑scale modes $k_H = 5.377 \times 10^{12}$ and $5.632 \times 10^{13}$ (see Figs. \ref{enumberL} (a), (b), (d)), the WKB approximation remains valid until $|x|$ approaches zero. For the large‑scale modes $k_H = 5.372 \times 10^{3}$ and $2.493 \times 10^{7}$, the WKB approximation becomes less accurate as the time approaches zero, nevertheless, it remains valid for a long time after the formation. For $|\epsilon_4/\lambda_j|$, $|\epsilon_7/\lambda_j|$ and $|\epsilon_8/\lambda_j|$ (see Figs. \ref{enumberL} (c), (e), (f)), one observes several bumps in the curves. These arise because the corresponding modes become gravitationally unstable. 
In particular, in Figs.~\ref{enumberL} (e) and (f),  near the bumps (unstable points) the WKB approximation fails ($|\epsilon_i/\lambda_j| > 1$), 
as discussed in Refs.~\cite{2jeans_static,2jeans_cosmo,2jeans_short}. 
The breakdown occurs because the Jeans scale in Eq.~\eqref{jeanss2t} yields  $\lambda_i = 0$ at these times. Note that in Figs.~\ref{enumberL}(a)–(f), $\lambda_i$ is not exactly zero near the bumps due to numerical accuracy. These bumps mark the times when a system becomes unstable.
Therefore, Eq.~\eqref{jeanss2t} is reasonably defined as the Jeans scale at the leading order.

\section{Sound Wave-Front in Two-Fluid Model}\label{Appa}

In this Appendix, we derive the sound wave-front for two-fluid model. Plugging Eq. \eqref{css} into Eq. \eqref{soundrqu} gives 
\bl
\frac{d\chi}{ d(\eta/L)}&=\frac{1}{\sqrt{\frac{9}{4}\frac{A}{A_\text{eq}}+3}}.\label{equationofsoundwave}
\el
Using Eq.\eqref{hduweifiu} in the above yields
\bl
d\chi&=\frac{(\eta_\text{eq}/L) d\eta'}{\sqrt{\frac{9}{2}+\frac{3\eta_\text{eq}^2}{L^2}-\frac{9}{2}\cos \eta'
+\frac{9}{2}\frac{\eta_\text{eq}}{L} \sin \eta'}},\label{dfhuirhuir}
\el
where the variable $\eta'\equiv\eta/L$ has been introduced.
We employ the integral identity 
\bl
\int\frac{d\phi}{\sqrt{a+b\cos\phi+c\sin\phi}}=
2\int\frac{d\psi}{\sqrt{a-p+2p\cos^2\psi}},\label{intgration}
\el
with the standard substitutions
$
\phi=2\psi+\alpha$,
$
\alpha=\tan^{-1}[\frac{b}{\sqrt{b^2+c^2}},\frac{c}{\sqrt{b^2+c^2}}]$,
$p=\sqrt{b^2+c^2}$ (see formula (2.580) in 
Ref.\cite{Intrgation}). 
Applying this transformation to \eqref{dfhuirhuir} gvies 
\bl
\int_{0}^{\chi}d\chi&=2\frac{\eta_\text{eq}}{L}\int^{\eta''}_{\eta''_i}\frac{d\eta''}{\sqrt{\frac{9}{2}+\frac{3\eta_\text{eq}^2}{L^2}-
\frac{9}{2}\sqrt{1+\frac{\eta_\text{eq}^2}{L^2}}
+9\sqrt{1+\frac{\eta_\text{eq}^2}{L^2}}\cos^ 2\eta''}},\label{fruiefhui}
\el
where we defined 
$\eta_i''\equiv-\frac12\frac{\eta_\text{max}}{L}$ and $\eta''\equiv\frac12\eta'-\frac12\frac{\eta_\text{max}}{L}$. 
Finally, using the elliptic integral identity
\bl
\int\frac{d\psi}{a-p+2p\cos^2\psi}=\frac{1}{\sqrt{a+p}}F\left[\psi,\frac{2p}{a+p}\right],
\el
where $F(a|b)$ is the elliptic integral of the first kind, and applying it to \eqref{fruiefhui} yields the closed form for the sound wave-front given in Eq. \eqref{newfhuirewfheuirnew}.

\end{CJK}
\end{document}